\documentclass[useAMS,usenatbib]{mn2e}

\usepackage[dvips]{graphicx}
\usepackage{epsfig}
\usepackage{amsmath}
\usepackage{subfigure} 
\usepackage{hyperref}
\usepackage{amssymb}
\title[Signatures of minor mergers in the Galactic disc] {Signatures of
  minor mergers in the Milky Way disc I: The SEGUE stellar sample}

\author[F.~A. G\'omez et al.]{Facundo A. G\'omez$^{1,2}$\thanks{Email:fgomez@pa.msu.edu},  
Ivan Minchev$^{3}$, 
\vspace{0.05cm}
Brian W. O'Shea$^{1,2,4}$, \newauthor 
Young Sun Lee$^{1,5}$,
Timothy C. Beers$^{1,6}$,
Deokkeun An$^{7}$,
James S. Bullock$^{8}$,\newauthor
Chris W. Purcell$^{8,9}$,
\'Alvaro Villalobos$^{10}$
\vspace{0.2cm}
\\
$^{1}$ Department of Physics and Astronomy, Michigan State University, East Lansing, MI 48824, USA\\
$^{2}$ Institute for Cyber-Enabled Research, Michigan State University, East Lansing, MI 48824, USA\\
$^{3}$ Leibniz-Institut f\"{ur} Astrophysik Potsdam (AIP), An der Sternwarte 16, D-14482, Potsdam, Germany\\
$^{4}$ Lyman Briggs College, Michigan State University, East Lansing, MI 48825, USA\\
$^{5}$ Joint Institute for Nuclear Astrophysics (JINA), Michigan State
University, East Lansing, MI 48824, USA \\
$^{6}$ National Optical Astronomy Observatory, Tucson, AZ 85719, USA\\
$^{7}$ Department of Science Education, Ewha Womans University, Seoul 120-750, Republic of Korea\\
$^{8}$ Center for Cosmology, Department of Physics and Astronomy, The University of California, Irvine\\
$^{9}$ Department of Physics and Astronomy, The University of Pittsburgh\\
$^{10}$ INAF-Osservatorio Astronomico di Trieste, Via Tiepolo 11,I-34143 Trieste, Italy
}

\begin{document}

\date{}

\pagerange{\pageref{firstpage}--\pageref{lastpage}} \pubyear{}

\maketitle

\label{firstpage}

\begin{abstract}

  It is now known that minor mergers are capable of creating structure
  in  the phase-space  distribution of  their host  galaxy's  disc. In
  order to search  for such imprints in the Milky  Way, we analyse the
  SEGUE F/G-dwarf and the Schuster  et al.  (2006) stellar samples. We
  find similar  features in  these two completely  independent stellar
  samples, consistent with the predictions of a Milky Way minor-merger
  event. We next apply the same analyses to high-resolution, idealised
  N-body simulations of the  interaction between the Sagittarius dwarf
  galaxy  and the  Milky  Way.  The  energy  distributions of  stellar
  particle samples  in small spatial  regions in the host  disc reveal
  strong variations  of structure with position. We  find good matches
  to  the observations  for models  with a  mass of  Sagittarius' dark
  matter halo progenitor  $\lessapprox 10^{11}$ M$_{\odot}$.  Thus, we
  show  that  this   kind  of  analysis  could  be   used  to  provide
  unprecedentedly   tight    constraints   on   Sagittarius'   orbital
  parameters, as well as place a lower limit on its mass.

\end{abstract}


\section{Introduction}

It is  now widely  accepted that the  gravitational pull exerted  by a
merging satellite galaxy can generate structure in the stellar disc of
its more-massive host. A striking example of the outcome of these kind
of  interactions is  the  system  composed of  M51  and its  companion
galaxy, NGC 5195 \citep[e.g.][]{oh,dobbs}. Previous studies have shown
that prominent spiral structure  is more commonly observed in galaxies
located within groups  or with companions \citep{kor,elma,elmb}. These
observations highlighted  the importance of tidal  interactions on the
structure of  galactic discs. Although  grand design spirals,  such as
the ones observed  in M51, are related to  massive companions (in this
case,  $M_{\rm  M51}  /  M_{\rm  NGC~  5195}  \gtrsim  0.3$),  several
theoretical studies  have shown that much less  massive satellites can
also    create    phase-space     structure    in    galactic    discs
\citep{qh93,tutu,kaza08,vh08,younger,quill09,min09,bird,
  gm12}. Morphological  features, such as spiral  arms, warps, flares,
central bars,  and low surface  brightness ringlike structures  in the
outskirts of  a disc  can be  the result of  either recent  or ongoing
accretion events.

Most  of  the previously-mentioned  morphological  features have  been
observed      in     the      disc     of      our      own     Galaxy
\citep[][]{levine,moma,cabre,pohl,new02,    yanny03},    and   various
formation  scenarios  have  been   explored  for  most  of  them.   An
interesting  example is the  Monoceros ring,  which is  a low-latitude
stellar structure  spanning about $180^{\rm o}$  in Galactic longitude
at nearly constant Galactocentric  distance. It has been proposed that
this structure  could be the remnant of  a tidally-disrupted satellite
galaxy   \citep{hnm,ibata03,conn,pena05}.    More  recently,   several
authors  have argued  that this  dynamically  cold ring  could be  the
result of a tidal perturbation from a satellite galaxy on the Galactic
disc \citep{kaza08,  younger,quill09}. In particular,  the Sagittarius
dwarf galaxy  (Sgr) has been proposed  as a possible  perturber of the
Milky  Way, causing the  emergence of  the Monoceros  ring as  the far
extension of a Galactic  spiral arm \citep{pur11} and also potentially
as a gravitational influence that circularised the orbit of a putative
dwarf galaxy that could have been  torn apart to become that same ring
\citep{md11}.  Motivated  by recent studies  on cosmological abundance
matching  \citep{cw09,bcw}, \citet[][hereafter  P11]{pur11} considered
models of Sagittarius with total masses as large as $\sim 10\%$ of the
Milky Way's  mass.  These models  can successfully reproduce  not only
dynamically cold structures such as the Monoceros ring, but also other
global morphological features observed in the Milky Way.  However, its
impact on the phase-space structure of the Solar neighbourhood has not
yet been explored.

As  first shown  by \citet{min09},  the  energy kick  imparted by  the
gravitational potential of a satellite  as it crosses the plane of the
disc can strongly perturb the  velocity field of disc stars located in
local  volumes such as  the Solar  neighbourhood.  In  stellar samples
close to the Sun (i.e.  distances $\leq 0.2$ kpc), these perturbations
can be observed in the $u$-$v$ plane as arc-like features traveling in
the direction  of positive $v$, where  $u$ and $v$ are  the radial and
tangential velocity  components, respectively.  In  a follow-up study,
\citet{gm12} showed that  the space defined by the  energy and angular
momentum  of  stars  is  a  better  choice  than  velocity  space,  as
substructure   remains    visible   even   in    much   larger   local
volumes. Furthermore, they also  showed that satellites with masses as
small as  $10\%$ of  the mass of  the host  can leave imprints  in the
phase-space  distribution  of  Solar neighbourhood-like  volumes  that
could  be  identified as  late  as  $\approx  5$~Gyr after  the  first
satellite's  pericentre passage.   Substructure  associated with  this
mechanism, known as  ``ringing,'' is expected to be  better defined in
the Galactic thick disc. This  is because these stars spend relatively
little  time near  the Galactic  plane, where  perturbations  from the
Galactic  bar, spiral structure,  and giant  molecular clouds  are more
vigorous.   Moreover,  the  thick  disc  is composed  of  a  very  old
population  of stars,  with ages  around  10 to  12 Gyr  \citep{sch06,
  bens07}. Thus, most of its stars must have been in place at the time
the hypothetical merger occurred.

An  obvious question  arises  from our  previous  discussion: Could  a
satellite  galaxy like  Sagittarius have  left imprints  of  its tidal
interaction  with the  Milky Way's  thick  disc in  the local  stellar
phase-space  distribution? Previous attempts  to describe  features in
the  velocity field  of the  Solar neighbourhood  have  been primarily
concerned with the dynamical  effects induced by non-axysimmetric disc
components,  such as  a central  bar or  self-gravitating  spiral arms
\citep[see,        e.g.,][]{walter,fux01,min07,min08,ant09,min10,ant11,
  quill11}.  These studies have  focused their attention on very small
local  volumes, dominated  mostly by  populations of  thin-disc stars.
However,  little attention  has been  paid to  the  identification and
characterisation of  substructure in local volumes of  the Milky Way's
thick  disc  that may  have  originated as  a  response  of the  tidal
interaction with a satellite galaxy \citep[see, e.g.,][]{min09}.

In  this work  we attempt  to fill  this gap  by analysing  a  full 6D
phase-space catalog  of F/G-type dwarf stars from  the Sloan Extension
for  Galactic  Understanding  and  Exploration (SEGUE;  Yanny  et  al.
2009).  The   great  advantage  of  this  catalog,   as  presented  by
\citet[][hereafter L11]{lee11b}, is that it contains accurate estimates
of $[\alpha/{\rm Fe}]$  and [Fe/H] ratios for a  large fraction of its
stars.  This enables  a chemical  separation of  the disc  system into
likely  thin-  and thick-disc  populations,  avoiding unwanted  biases
associated  with  methods  based  on  stellar  kinematics  or  spatial
distributions.   To  complement this  analysis,  we  also explore  the
distribution of disc stars from the \citet{sch06} catalog.

The paper's outline  is as follows. Section 2  presents a brief review
of  the main  properties  of ringing.  In  Section 3  we describe  the
stellar samples  analysed in this  work. The distribution of  stars in
energy  and   angular  momentum  space   of  the  F/G-dwarf   and  the
\citet{sch06}   catalogs   are   analysed   in  Section   4   and   5,
respectively.   In  Section   6  we   apply  the   same   analyses  to
high-resolution   N-body  simulations   of  the   interaction  between
Sagittarius and the Milky Way. A summary and conclusions are presented
in Section 7.

\section{Ringing in galactic discs}
\label{sec:model}

\begin{figure}

\hspace{-0.1cm}
\includegraphics[width=80mm,clip]{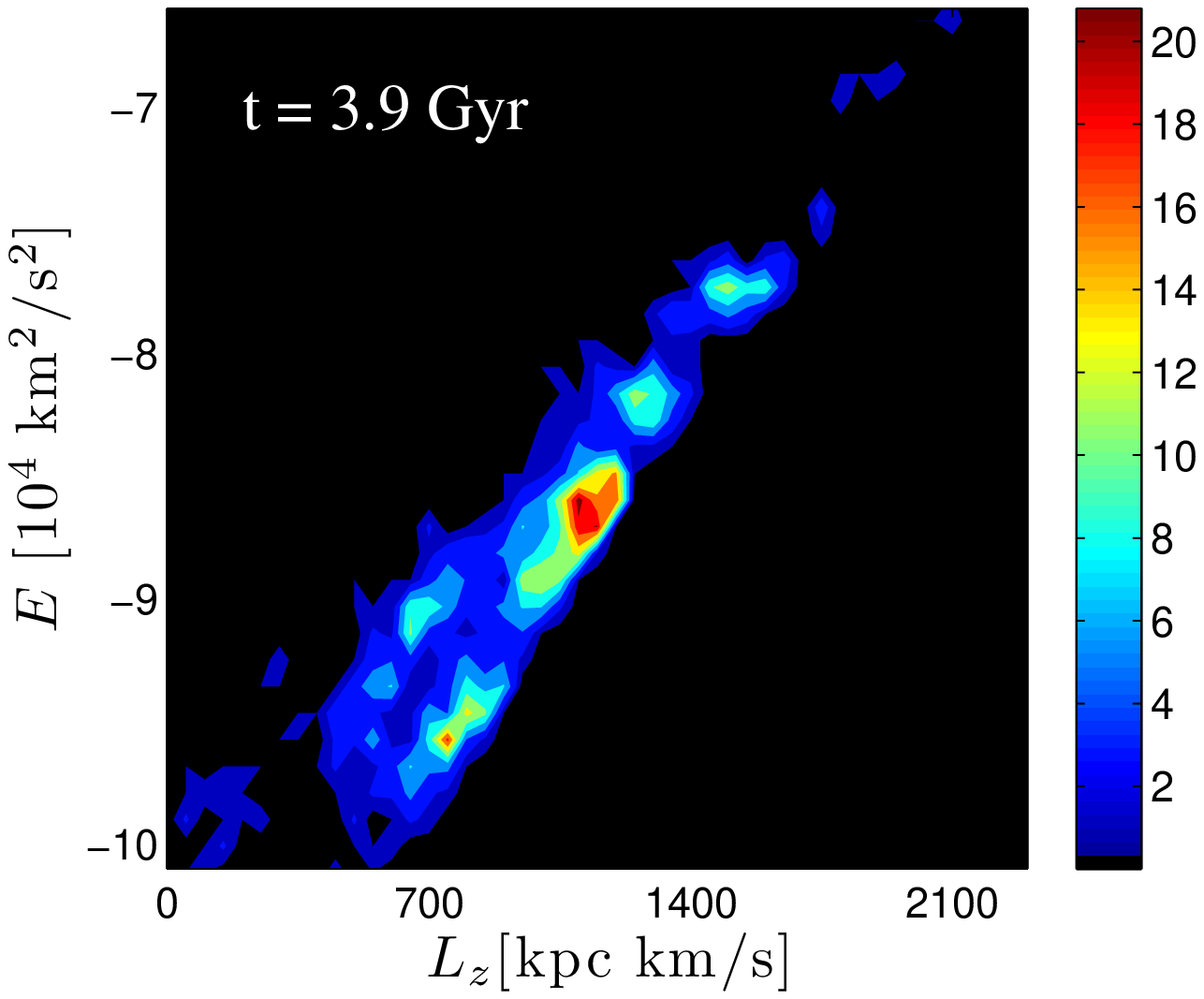}
\\
\includegraphics[width=69mm,clip]{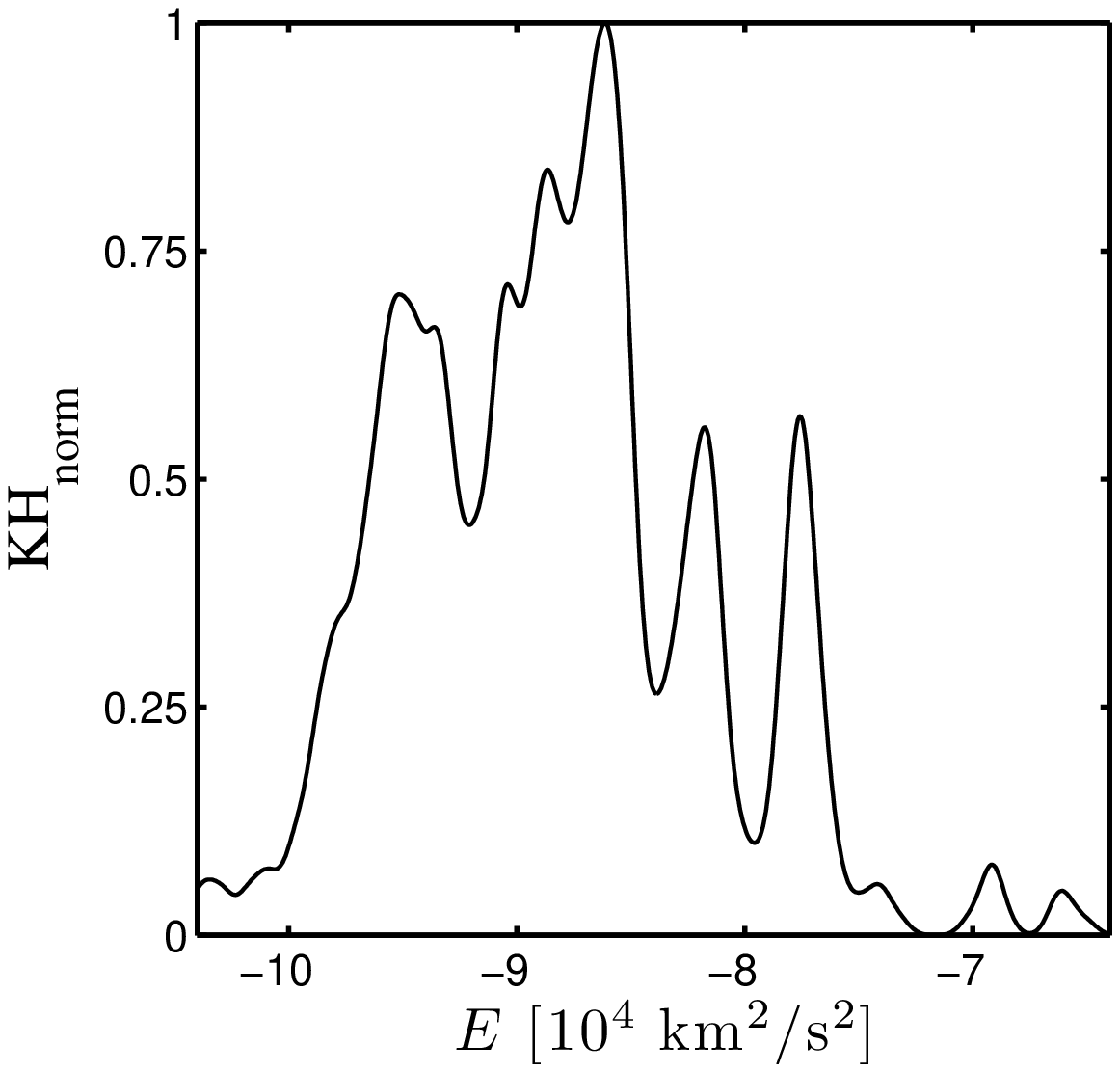}
\caption{Top panel: $E-L_{z}$ distribution  of disc particles inside a
  sphere of 2.5~kpc radius located  at 8~kpc from the galactic centre,
  obtained from a  simulation with a mass ratio  of $M_{\rm host}^{\rm
    total}  / M_{\rm  sat}^{\rm total}  =  0.2$. The  total number  of
  particles enclosed within this volume is $N_{\rm part}^{\rm model} =
  976$.   The snapshot  corresponds to  a  time of  3.9~Gyr after  the
  satellite  was launched,  or to  1.9~Gyr after  the  satellite fully
  merged  with  the  host  galaxy. The  different  colours  (contours)
  indicate  different number of  particles. Note  the large  number of
  lumps with  nearly constant  energies associated with  density waves
  crossing  the  sphere.   Bottom  panel:  Kernel  histogram  (KH)  of
  energies,  $E$, obtained  from the  same distribution  of particles.
  Each  peak in  the kernel  histogram corresponds  to a  density wave
  observed in the top panel.}
\label{fig:model}
\end{figure}

\citet{min09}, followed by  \citet[][hereafter G12]{gm12}, showed that
relatively  massive minor-merger events  can generate  substructure in
the   velocity   field   of   disc   stars  located   in   the   Solar
neighbourhood. Merging  satellite galaxies with total  masses as small
as $10\%$ of the mass of  the Milky Way may excite density waves whose
signatures can be identified  even $5$~Gyr after the first satellite's
pericentre  passage.    The  space  of   $E$-$L_{z}$  is  particularly
well-suited to identify density waves, since the waves are surfaces of
constant  energy  \citep[see  also e.g.][]{hz00,hwzz99}.   Over  time,
these  constant-energy  features  associated  with the  density  waves
become  more   closely  spaced,  and  their  number   increases  as  a
consequence  of phase wrapping  (see Section  3 of  G12).  In  the top
panel  of Figure~\ref{fig:model}  we show,  in $E$-$L_{z}$  space, the
distribution    of   disc   particles    located   within    a   Solar
neighbourhood-like sphere of 2.5 kpc  radius, obtained from one of the
simulations analysed in  G12. The sphere is centred  at 8~kpc from the
galactic  centre and the  distribution was  obtained after  3.9~Gyr of
evolution, which corresponds  to a time since total  disruption of the
satellite galaxy of 1.9 Gyr.  In this simulation, the satellite galaxy
has a total initial mass equal to $20\%$ of the mass of the host. Note
that only $N_{\rm part}^{\rm model} = 976$ are located within this 2.5
kpc  sphere.  It  is  clear  that disc  particles  are distributed  in
different  lumps of nearly  constant energy,  each of  them associated
with  a   different  density-wave  crossing.   The   bottom  panel  of
Figure~\ref{fig:model}  shows  a Kernel  Histogram  (KH) of  energies,
$E$. To compute the kernel histogram, a Gaussian kernel with bandwidth
$\sigma = 0.011~\Delta E$ was assigned to each particle. Here, $\Delta
E = 4 \times 10^{4} {\rm ~km^{2} ~s^{-2}} $ is approximately the total
extent  of the  satellite in  energy  space within  this volume.   The
kernel histogram  is the  sum of all  of the Gaussian  kernels.  Since
density  waves  are  surfaces  of  nearly constant  energy,  they  are
observed in the KH as  well-defined peaks.  Note that larger values of
$\sigma$  tend  to  oversmooth  the  KH, erasing  some  of  the  peaks
associated with substructure observed in $E-L_{z}$ space.  Conversely,
smaller  values produce  noisier KHs.   In  what follows  we keep  the
values of $\sigma$ and $\Delta E$ fixed at these values.

\section{The stellar samples}

\subsection{The SEGUE F/G-Dwarf sample}
\label{sec:gdwarf}

The SEGUE F/G-dwarf sample analysed in this work was culled from $\sim
~70,000$ F and G type stars with available low-resolution spectroscopy
($R \sim  2000$), as provided  in SDSS DR8 \citep{aiha}.   Among them,
about $63,000$ stars  were targeted as G-dwarfs; hence,  our sample is
dominated  by  the  G-dwarf  candidates  used  by  L11.   The  G-dwarf
candidates  were   obtained  by  selecting  stars   with  colours  and
magnitudes in the range $0.48 <  (g-r)_{0} < 0.55$ and $r_{0} < 20.2$,
respectively, while the  rest of the sample covers  $0.2 < (g-r)_{0} <
0.48$, and $g_{0}  < 20.2$. Thanks to this  simple selection function,
the stellar sample is expected  to be completely unbiased with respect
to  kinematics, and  only  slightly biased  towards  metal poor  stars
\citep[see][]{katie}.

\subsubsection{Local sample}

In order to obtain a  local sample of stars with accurate measurements
of  their 6D  phase-space coordinates,  as well  as  metallicities and
$[\alpha/{\rm  Fe}]$ ratios,  a series  of  cuts were  applied to  the
complete catalog. Here we provide a brief summary of the criteria used
for the  sample selection, but we  refer the interested  reader to L11
for a more  detailed description of this procedure  as we follow their
prescription to select the final sample.

Stellar atmospheric parameters, such as effective temperature, $T_{\rm
  eff}$, surface gravity, $\log~g$, and metallicity, $[{\rm Fe}/{\rm
  H}]$, were determined using the SEGUE Stellar Parameter Pipeline,
SSPP \citep{lee08a,lee08b,ap08,smo}.  In order to obtain high-quality
estimates of both $[{\rm Fe}/{\rm H}]$ and $[\alpha/{\rm Fe}]$
\citep{lee11a}, only stars with spectra of signal-to-noise ratios
(S/N) greater than 30 ${\rm \AA}^{-1}$ were considered.  Importantly,
this also ensures that errors in the estimated radial velocities are
smaller than 5 km s$^{-1}$. Proper motions were obtained following
\citet{munn04}, after correcting for the systematic error described in
\citet{munn08}.  Distances to individual stars were estimated using
calibrated set of stellar isochrones \citep{an09b}, following the
prescription of \citet{an09a}.  To minimise possible distance bias
from stellar age effects near the main sequence turnoff, only stars
with $\log~g \geq 4.2$ were considered. In addition, to minimise the
errors on the estimates of phase-space coordinates, we only consider
stars with distances from the Sun $0.4 \lesssim d \leq 2$ kpc. Note
that the inner limit is imposed by the data itself, due to saturation
of the SDSS photometric scans for $g \lesssim 14.5$.

To  compute  Galactocentric  positions  and velocities  of  our  local
stellar  sample, we  assume  the Sun  to  be located  at $R_{\odot}  =
8$~kpc, and that it has a  peculiar velocity with respect to the Local
Standard  of  Rest  (LSR)  $(U,~V,  ~W)_{\odot}  =  (11.1,~12.2,~7.3)$
km~s$^{-1}$ \citep{scho10}.  We further assume  a velocity of  the LSR
with  respect  to   the  Galactic  centre  of  $V_{\rm   LSR}  =  220$
km~s$^{-1}$. Finally,  we discard stars with $[{\rm  Fe}/{\rm H}] \leq
-1.2$ and $V_{\phi}  < 0$ km s$^{-1}$ in  order to avoid contamination
(as much as  possible) from the stellar halo, as  well as the proposed
metal-weak thick-disc component of the Galaxy \citep{carollo10}.

\subsubsection{The thin- and thick-disc populations}
\label{sec:thin_thick}

As mentioned  in the Introduction, signatures of  ringing are expected
to be stronger in the thick disc than the thin disc because its stars
are   less  affected   by   perturbations  from   the  Galactic   bar,
self-gravitating  spiral  arms,   and  giant  molecular  clouds.  More
importantly, the thick  disc is mainly composed of  a very old stellar
population, with ages of $\sim 10-12$~Gyr \citep{sch06, bens07}. Thus,
most of its stars must have been  in place as recently as 5~Gyr ago --
a required condition  if, for example, we seek  to identify signatures
of a merger event that may have started approximately at that time.

Following L11, we split our local sample into likely thin- and thick-disc
populations on the basis of their stellar chemical abundances, i.e.,
$[\alpha/{\rm Fe}]$ and [Fe/H]. We assign stars to the thin-disc
(low-$[\alpha/{\rm Fe}]$) and thick-disc (high-$[\alpha/{\rm Fe}]$)
components according to the following scheme:

\begin{description}

\item[For stars] with [Fe/H] $\geq -0.8$

\begin{itemize}
\item thin disc, if $[\alpha/{\rm Fe}] < -0.08 ~ \cdot$ [Fe/H] $+ 0.15$
\item thick disc, if $[\alpha/{\rm Fe}] > -0.08 ~ \cdot$ [Fe/H] $+ 0.25$
\end{itemize}

\item[For stars] with [Fe/H] $< -0.8$

\begin{itemize}
\item thin disc, if $[\alpha/{\rm Fe}] < +0.214$
\item thick disc, if $[\alpha/{\rm Fe}] > +0.314$
\end{itemize}

\end{description}

While any old stellar population with high velocity dispersion would be
suitable for our investigation, by assigning stars to the different
components of the disc according to $[\alpha/{\rm Fe}]$ we avoid
introducing unwanted biases associated with methods based on stellar
kinematics or spatial distributions. Note the gap of 0.1~dex left in the
$[\alpha/{\rm Fe}]$ cuts between the thin disc and thick disc, which is
chosen to avoid misclassification of stars. For a detailed description of
the motivation behind this scheme, we refer the reader to section 3.1 of
L11.

\subsection{The Schuster et al. (2006) sample}
\label{sec:schuster}

The \citet[][hereafter SCH06]{sch06} catalog comprises a total of 1533
high-velocity and metal-poor stars. Positions and proper motions for these
stars were primarily derived from Hipparcos \citep{hip}, Tycho-2
\citep{tycho}, and the revised NLTT catalog \citep{sg03}, whereas radial
velocities were obtained from a number of different sources available in
the literature (see SCH06, and references therein). As in Section
~\ref{sec:gdwarf}, we compute Galactocentric velocities by assuming
$R_{\odot} = 8$ kpc, $(U,~V,~W)_{\odot} = (11.1,~12.2,~7.3)$~km s$^{-1}$
and $V_{\rm LSR} = 220$ km s$^{-1}$.

Metallicities estimated  via a  photometric calibration based  on {\it
  uvby-$\beta$} photometry are also provided for the entire sample. As
shown by  SCH06, it is  possible to discriminate stars  from different
Galactic  components by means  of the  $X$ parameter,  where $X$  is a
simple linear  combination of the rotational  velocity and metallicity
of the  stars. Following SCH06, we  discard from our  sample all stars
with $X  \geq 0$, in order  to avoid potential  contamination from the
Galactic halo\footnote{Note that only a fraction smaller than $3\%$ of
  our final F/G-dwarf thick-disc subsample has $X \geq 0$.}.

Note that estimates of $[\alpha/{\rm Fe}]$ are not provided in the SCH06
catalog. Thus, a subdivision between likely thin- and thick-disc stars
cannot be performed following the scheme described in
Section~\ref{sec:thin_thick}. Nevertheless, as shown by \citet{ns91} and
SCH06, the thin-disc population in this sample belongs to a very old disc
component, with the bulk of the population older than 10~Gyr, and the
remaining stars with ages between 10 and 4~Gyr. We thus ensure that most of
these stars were already part of the disc $\sim 5$ Gyr ago. It is important
to note that all disc stars in this sample are located at a distance from
the Sun $d < 0. 2$~kpc. Thus, this sample probes the innermost regions of
the Solar neighbourhood, complementing the more distant SDSS F/G-dwarf sample.

\section{The SEGUE F/G-Dwarf sample in $E-L_{z}$ space}
\label{sec:gdwarf_ener}

We now analyse the $E-L_{z}$ distribution of the stars present in our local
SEGUE F/G-dwarf sample. In order to obtain an estimate of the orbital energy
of these stars we first need to assume a shape for the Galactic potential.
We adopt a model consisting of three different components. These are:
(1) a Miyamoto-Nagai disc
\citep{mi-na}
\begin{equation}
\Phi_{\rm disc}=-\frac{GM_{\rm d}}{\sqrt{R^{2}+(r_{a}+\sqrt{Z^{2}+r_{b}^{2}})^{2}}},
\end{equation}
(2) a Hernquist bulge \citep{hernq},
 \begin{equation}
\Phi_{\rm bulge}=-\frac{GM_{\rm b}}{r+r_{c}},
\end{equation} 
and (3) a NFW dark matter halo \citep{nfw} 
\begin{equation}
\Phi_{\rm halo}=-\frac{GM_{\rm vir}}{r\left[\log(1+c)-c/(1+c)\right]}\log\left(1+\frac{r}{r_{s}}\right).
\end{equation}
\begin{table}
\begin{minipage}{90mm} \centering
  \caption{Parameters of the present day Milky Way-like potential used
    in this work. Masses are in M$_{\odot}$ and distances are in kpc.}
\label{table:model}
\begin{tabular}{@{}lll} \hline Disc & Bulge & Halo\\
  \hline $M_{\rm d} = 4 \times 10^{10}$ & $M_{\rm b} = 8 \times 10^{9}$ & $M_{\rm vir} = 1 \times 10^{12}$ \\ $r_{\rm a} = 6.5$ & $r_{\rm c} =0.7$ & $r_{s} = 21.5$ \\  $r_{\rm b} = 0.26$ & & $c = 12$ \\
  \hline
\end{tabular}
\end{minipage}
\end{table} 

Table~\ref{table:model}  summarises   the  numerical  values   of  the
parameters used for our Galactic potential model. These parameters are
similar  to those  used by  other  authors \citep[e.g.,][]{bj05,gea10,
  pena10}  and fit the  Milky Way  rotation curve  \citep{kly02}. Note
that,  as  described by  \citet[][and  references therein]{gh10},  the
results are  not strongly  dependent on the  particular choice  of the
potential, since substructure in  integrals of motion space are robust
to small differences in the  mass distribution because we always focus
on  small volumes  in space.  Hence,  small changes  in the  potential
essentially  act  as a  zero-point  offset,  affecting  all the  stars
present in this volume in the same way.
 
\begin{figure}
\hspace{-0.08cm}
\includegraphics[width=85mm,clip]{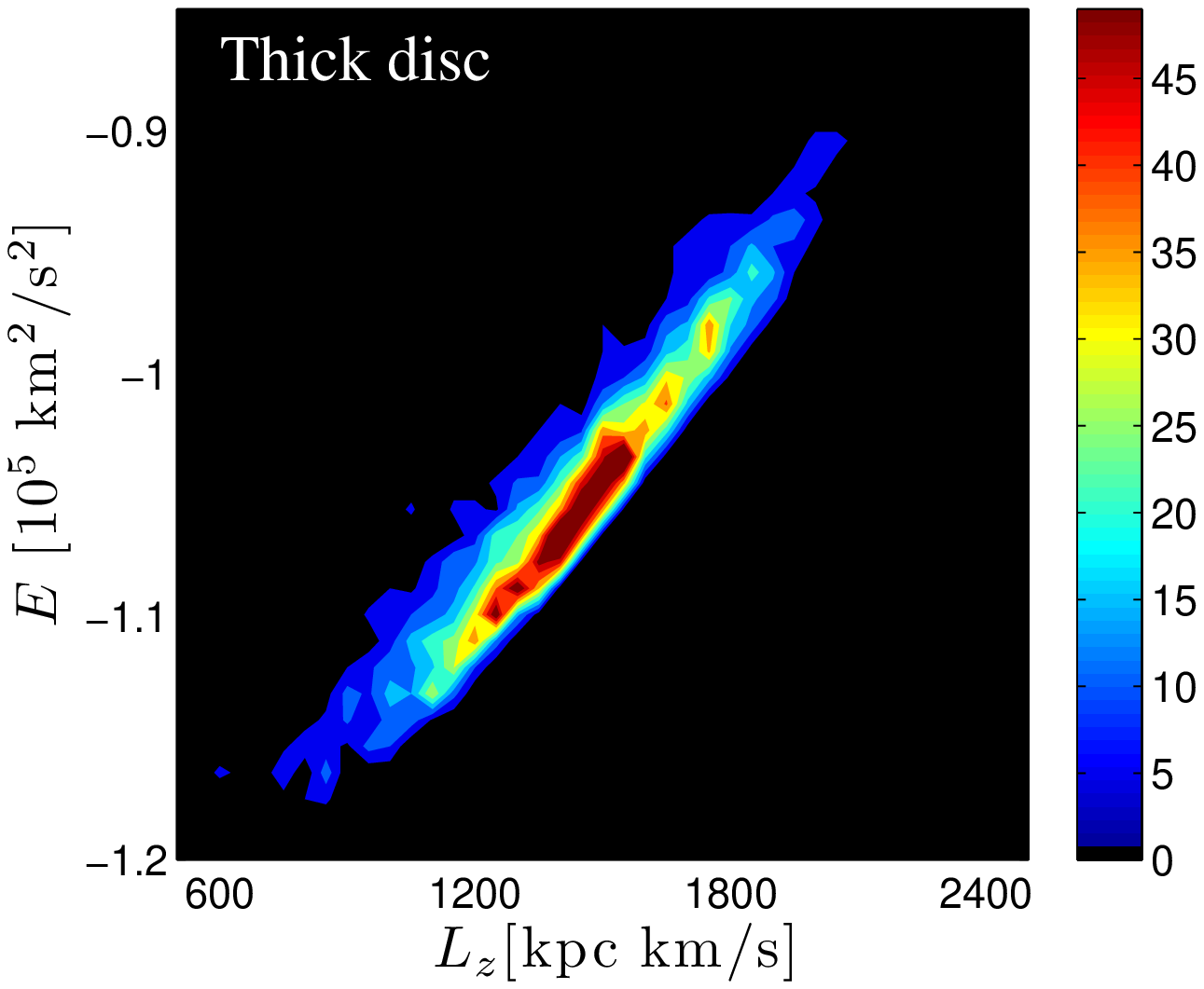}
\\
\includegraphics[width=72.5mm,clip]{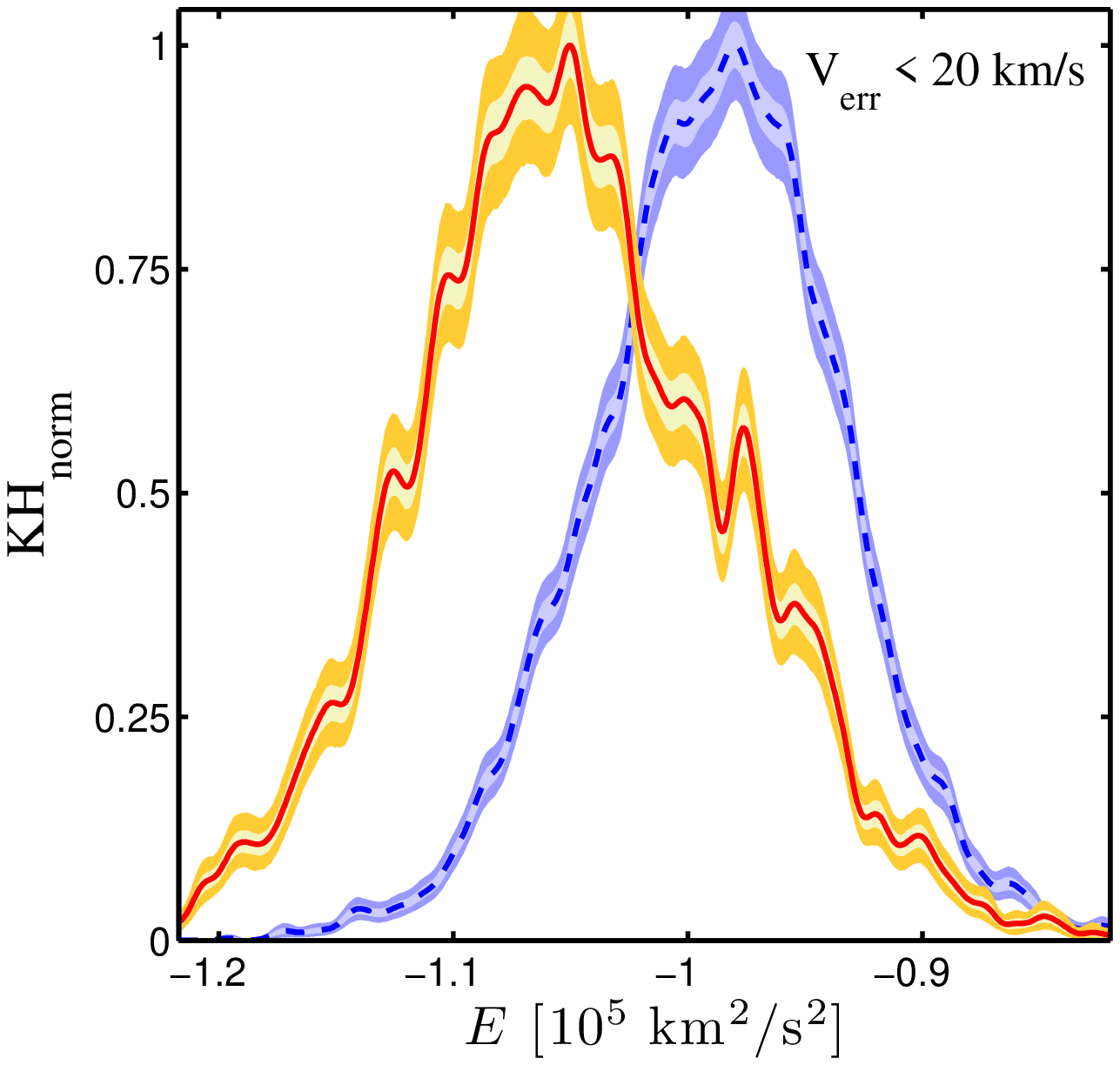}
\caption{Top panel: $E-L_{z}$  distribution of likely thick-disc stars
  located  within  2~kpc  of  the  Sun, obtained  from  our  F/G-dwarf
  sample. The  different colours (contours)  indicate different number
  of stars.  Only stars with total  errors in velocity  $V_{\rm err} <
  20$  km  s$^{-1}$ are  considered.  The  subsample contains  $N_{\rm
    stars}^{\rm  thick} = 3141$  high-$[\alpha/{\rm Fe}]$  stars. Note
  the  presence of a  series of  high-density features,  especially at
  large  values of energies.  Bottom panel:  kernel histogram  of $E$,
  obtained from both the thick-disc (red solid line) and the thin-disc
  (blue  dashed  line)  stellar  subsamples.  The  thin-disc  subsample
  contains $N_{\rm  stars}^{\rm thin} =  4903$ low-$[\alpha/{\rm Fe}]$
  stars.  The  shaded  areas  indicate  errors  associated  with  poor
  sampling   of   the  underlying   distributions   (see  text).   The
  high-density features observed in  the top panel correspond to peaks
  in the kernel histogram.  Note that the thin-disc subsample exhibits
  a smoother distribution relative to the thick disc. }
\label{fig:el_20}
\end{figure}

\subsection{Thin vs. Thick disc}
\label{sec:th_tk}

The  top panel  of  Figure~\ref{fig:el_20} shows  the distribution  of
$\alpha$-enhanced,  likely thick-disc  stars in  $E-L_{z}$  space. For
this figure we have only considered stars with $d \leq 2$ kpc from the
Sun  and a  total velocity  error $V_{err}  < 20$  km s$^{-1}$.   As a
result,  and because  of the  different cuts  applied to  the complete
catalog (see Section~\ref{sec:gdwarf}), we  are only left with a total
of  $N_{\rm stars}^{\rm  thick} =  3141$ likely  thick-disc  stars.  A
series  of  high-density  features  can  be observed  in  this  panel,
especially at  high values of  $E$ and $L_{z}$. These  features become
more evident in the kernel histogram of energies, shown as a red solid
line  in  the  bottom  panel  of the  same  figure.   Comparison  with
Figure~\ref{fig:model} clearly  indicates that, if  these features are
to be associated with ringing, they have a much smaller amplitude than
might be expected.  However, due  to the relatively large threshold in
total  velocity  errors  considered  (i.e.,  $V_{\rm  err}  <$  20  km
s$^{-1}$), these  peaks may  have been smoothed  out; we  explore this
issue further in Section 4.2. On the other hand, due to the relatively
low number  of stars, some of the  peaks may have arisen  due to noise
associated  with poor  sampling  of the  underlying distribution.   To
investigate  this, we  bootstrapped our  thick-disc  stellar subsample
2000  times  and computed  a  kernel histogram  for  each  one of  the
realisations. At each value of $E$,  we sort the results of the KHs in
ascending   order.   The  shaded   areas  in   the  bottom   panel  of
Figure~\ref{fig:el_20} show  the 25-75 percentiles  (light shadow) and
the  5-95 percentiles  (dark  shadow) of  the  distribution of  values
obtained  from the bootstrapped  KHs.  The  results of  this procedure
indicate that  only a few of  these peaks have  amplitudes with values
above  the noise  level. However,  in the  following sections  we show
that,  after accounting  for the  effects of  measurement  errors, and
through exploration of the SCH06  catalog, some of the remaining peaks
may indeed be real.

\begin{figure}
\includegraphics[width=80mm,clip]{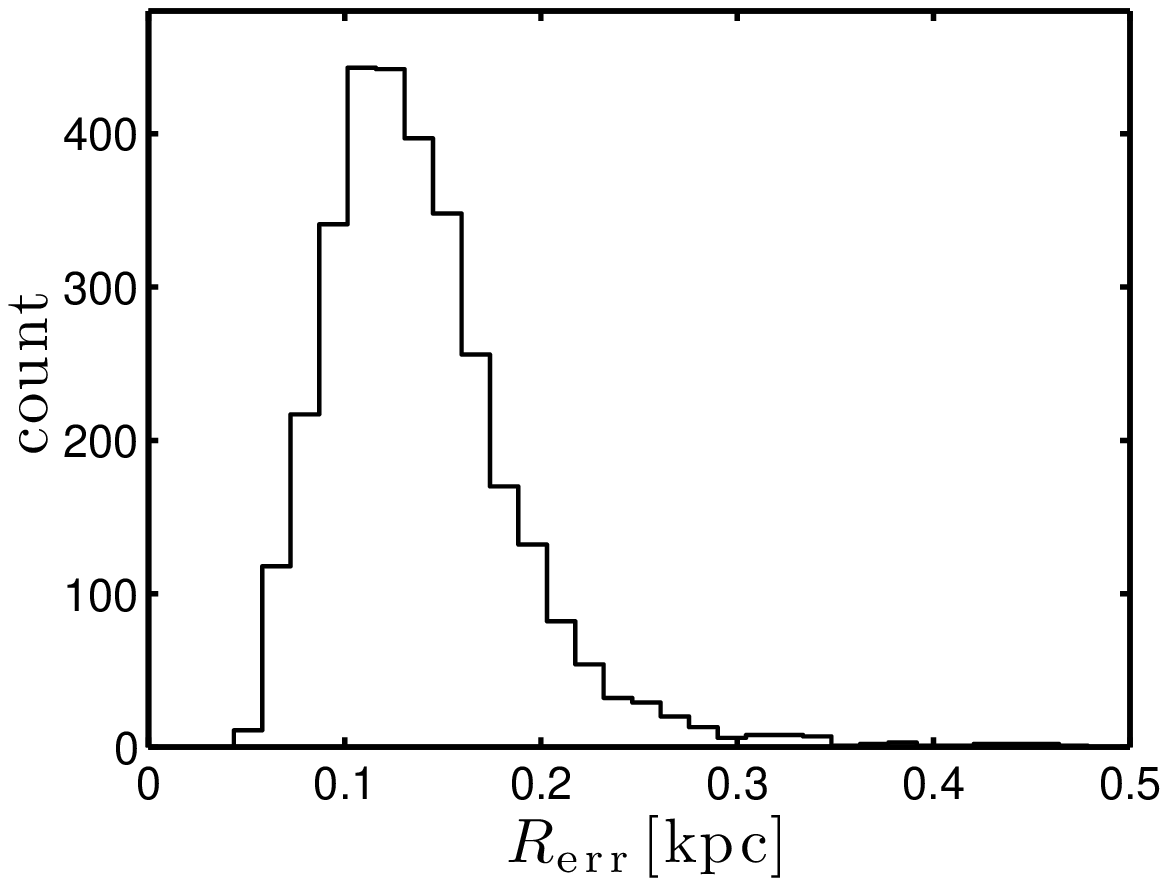}
\\
\includegraphics[width=80mm,clip]{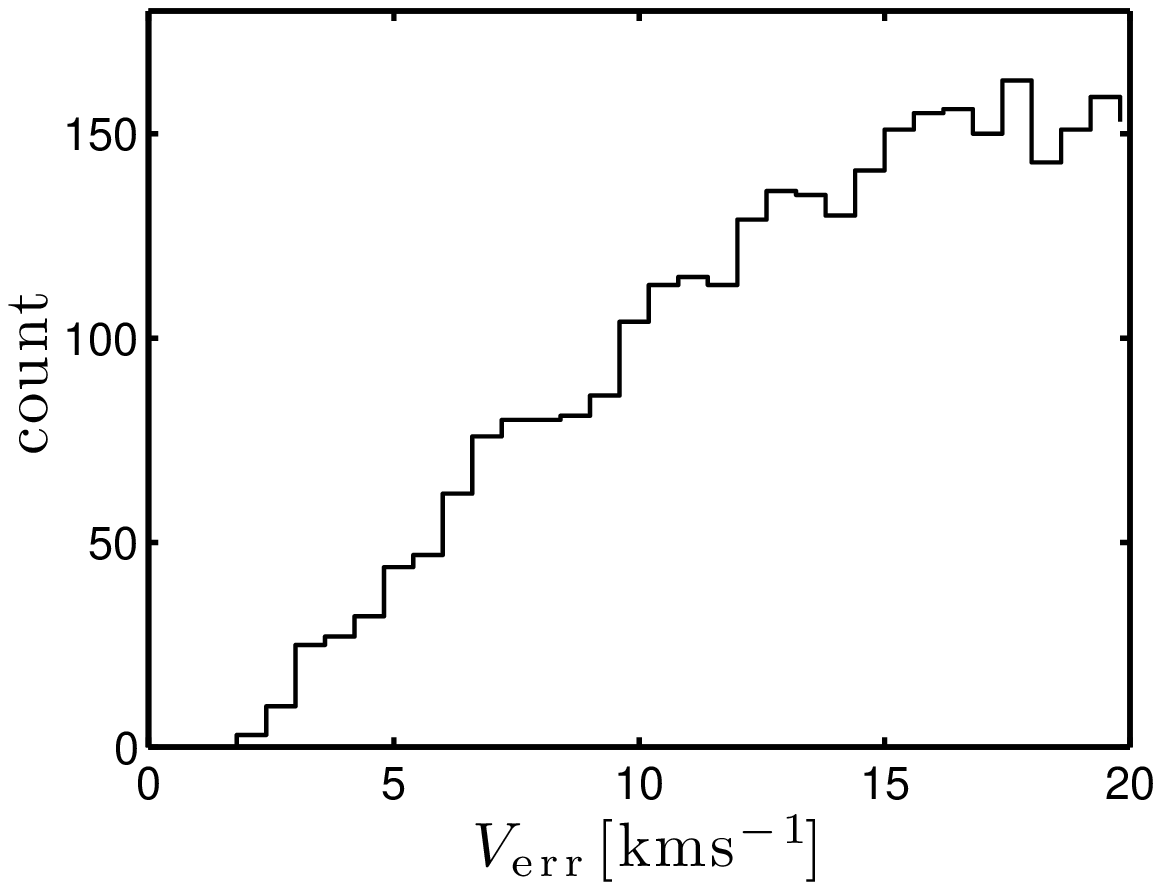}
\caption{Distribution of total errors  in positions (top panel) and
  velocities (bottom panel), extracted from our likely thick-disc SDSS F/G-dwarf
  subsample.}
\label{fig:errors}
\end{figure}

\begin{figure*}
\includegraphics[width=61.5mm,clip]{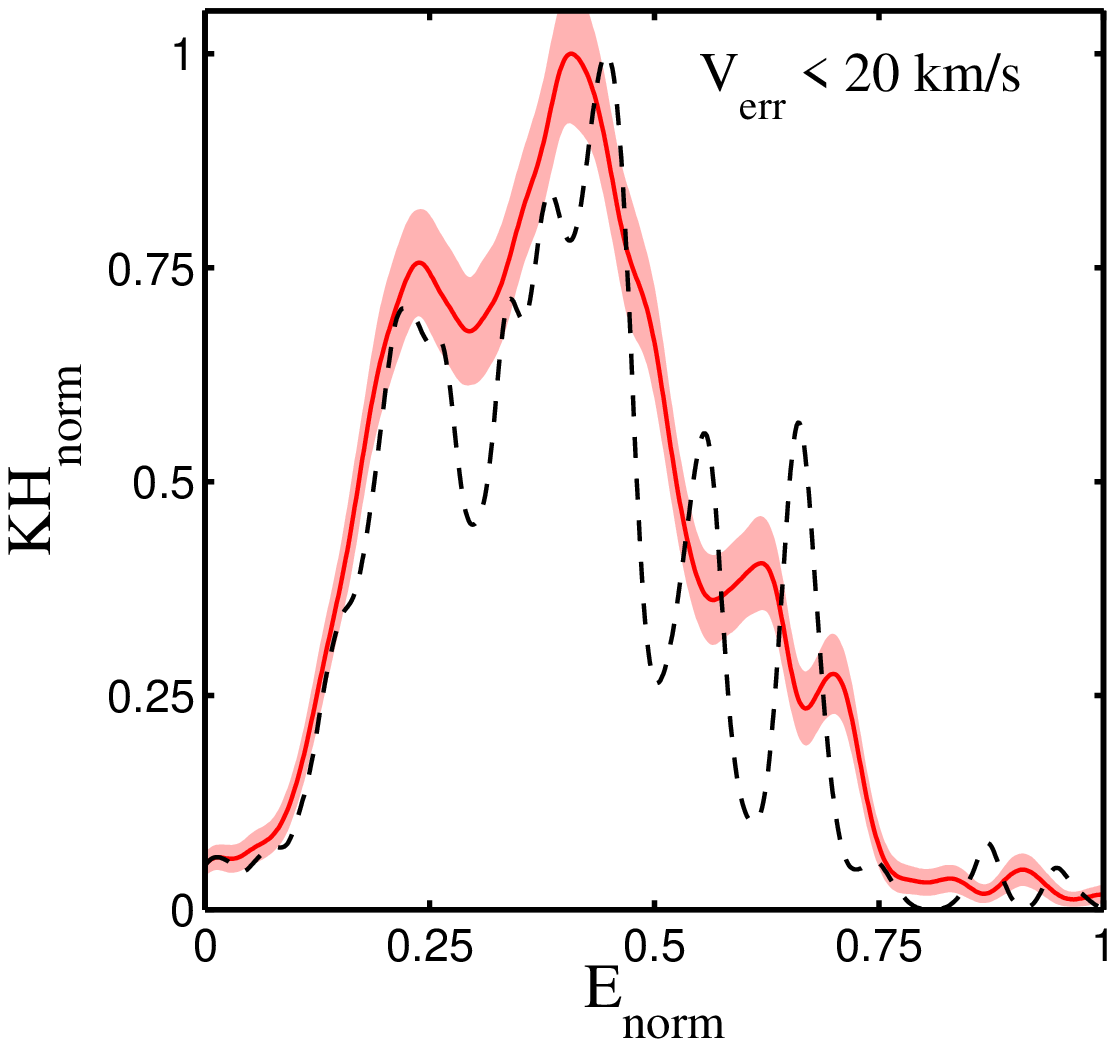}
\includegraphics[width=50mm,clip]{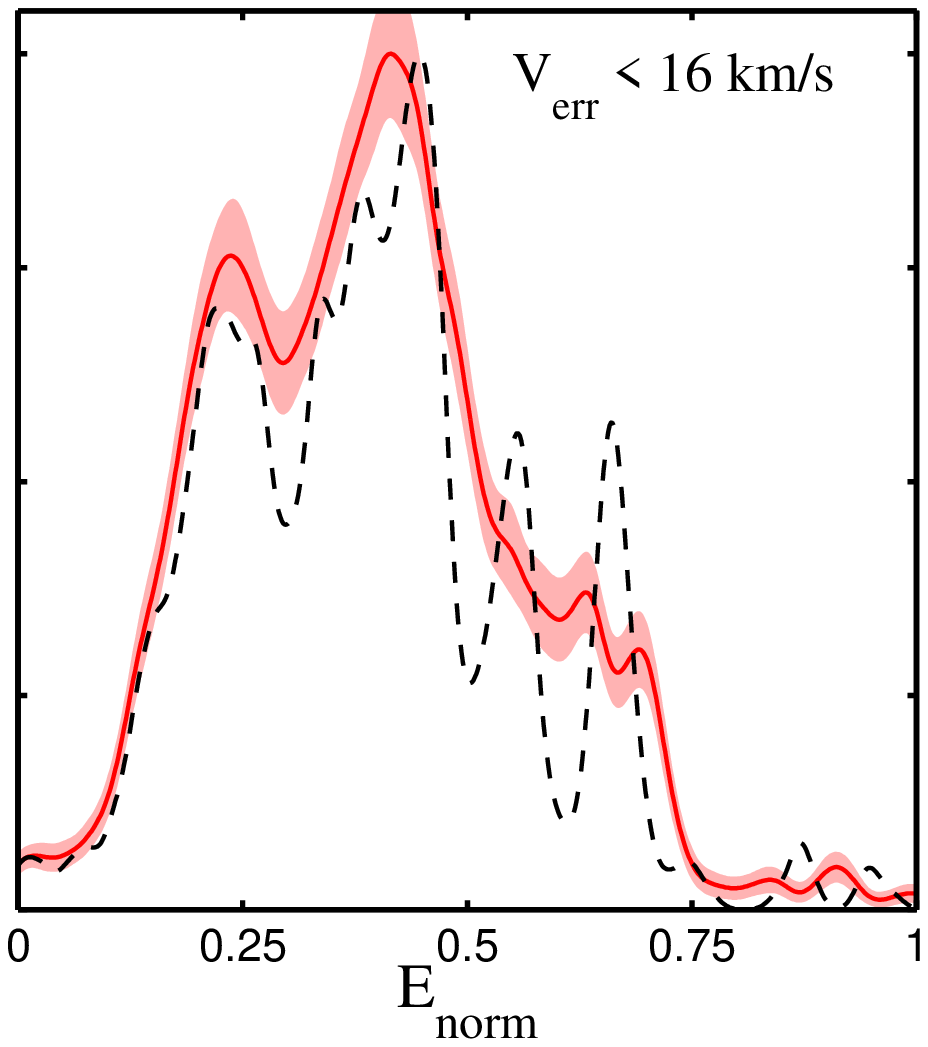}
\includegraphics[width=50mm,clip]{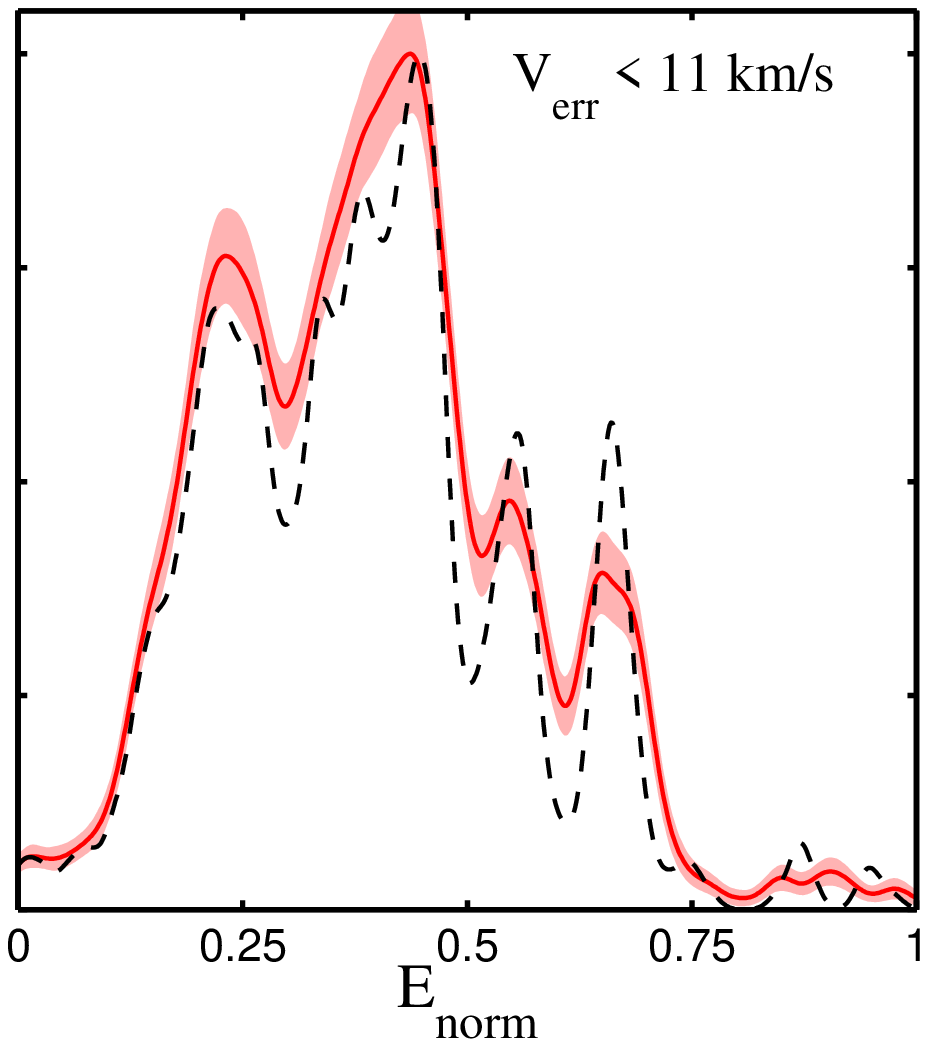}

\caption{Kernel histograms of normalised  energies, obtained  from the
  distribution of particles shown in Figure~\ref{fig:model}, after
  convolving with measurement errors. The red solid line in each panel
  shows the averaged kernel histogram obtained after 400 realisations of the error
  convolution. The maximum error in total velocity considered is indicated
  in the top right corner of each panel. The shaded regions indicate
  $1\sigma$ deviations from the mean. For comparison we show, with a dashed
  black line, the kernel histogram obtained from the same distribution of particles
  before convolution. When errors in total velocity as large as 20 km/s are
  considered, most of the peaks in the histogram are erased (left panel). These
  peaks are recovered when we reduce the magnitude of the maximum error to
  11 km/s (right panel).}
\label{fig:mod_errors}
\end{figure*}

The bottom panel of Figure~\ref{fig:el_20} also shows (blue dashed line) the
kernel histogram on $E$ of the stars associated with the low-$\alpha$, likely thin-disc
stars. In this case, the total number of stars in the subsample is $N_{\rm
stars}^{\rm thin} = 4903$. It is very interesting to observe that the
distributions associated with the thin disc and thick disc are shifted with
respect to one another. The thin-disc subsample comprises a population of
stars that, at least locally, is less tightly bound than that of the thick
disc. Note that, within volumes as small as the ones considered here, the
Galactic potential can be regarded as constant. Thus, for these stars, the
binding energy is mainly a function of their velocities. Since the thick
disc is composed of a population of stars with a lag in $V_{\phi}$ of $\sim
30$ km s$^{-1}$ relative to the thin disc (as shown in, e.g., L11), it is not surprising
to find that these stars have a larger binding energy\footnote{Note that
the mean radial and vertical velocity components for both populations are
$\sim 0$ km s$^{-1}$.}. It is also noticeable that the thick-disc stars
exhibit a much broader distribution in energy than their thin-disc counterparts. This
is a manifestation of the larger velocity dispersion (in all directions)
associated with the thick-disc component. Finally, note that the thin disc
exhibits a much smoother distribution when compared to that of the thick
disc. One has to bear in mind that the thin-disc
subsample is also $\sim 35\%$ larger than the thick-disc subsample. Thus,
noise associated with poor sampling is reduced. This is indicated by the
tighter blue and light-blue shaded regions, computed as discussed above.
Nevertheless, in Section~\ref{sec:errors} we show that this result holds
even when considering smaller stellar subsamples with more accurately
determined velocities.

\subsection{Characterising  the  effects   of  errors  on  phase-space
  coordinates}
\label{sec:errors}

As noted above, the  measurement errors in the phase-space coordinates
of  our local F/G-dwarf  stellar sample  may be  large enough  to either
smooth out  or erase  some of the  signatures of ringing  in $E-L_{z}$
space. To  characterise their effect  on the observed features  in the
kernel histograms, we have  extracted the distributions of errors from
our  thick-disc subsample, and  convolved them  with the  position and
velocities of the particles shown in Figure~\ref{fig:model}. Estimates
of  the errors were  obtained from  the SSPP  (see L11  and references
therein);       their      distributions       are       shown      in
Figure~\ref{fig:errors}. The left panel of Figure~\ref{fig:mod_errors}
shows the  results of  this convolution when  errors in  velocities as
large as  20 km$^{-1}$  are considered. The  red solid line  shows the
averaged  kernel  histogram of  normalised  energies, $E_{\rm  norm}$,
obtained after  400 different  realisations of the  error convolution,
whereas  the shaded  region  indicates its  standard deviation.  Here,
$E_{\rm norm} = (E_{i} - \min(E)) / \Delta E$, with $\Delta E$ defined
as  in Section~\ref{sec:model}.  It is  important to  remark  that, to
compute the energies of  the $N$-body particles after convolution with
errors, we have used the same analytic potential that was used for the
F/G-dwarf sample. Note that this  simulation was designed to emulate the
result of a minor merger experienced  by a Milky Way-like galaxy at $z
=  1$. Thus, the  properties of  this potential  were scaled  to those
expected at  $z=1$, as described  in Section 2 of  G12. Interestingly,
due  to the  large magnitude  of the  errors considered,  most  of the
structure associated  with ringing has been  erased. Furthermore, some
spurious peaks can  be observed.  For comparison we  show with a black
dashed line the  kernel histogram obtained from this  set of particles
before error  convolution. The middle  panel of this figure  shows the
results of  the same  procedure, now after  convolving with  errors in
velocity $V_{\rm  err} < 16$ km  s$^{-1}$. Note that  these errors are
still large  enough to erase most  of the signatures  of ringing. Only
when considering  errors in  velocity $V_{\rm err}  < 11$  km s$^{-1}$
(right panel) do we start recovering the previously-observed peaks.

\begin{figure}
\includegraphics[width=82mm,clip]{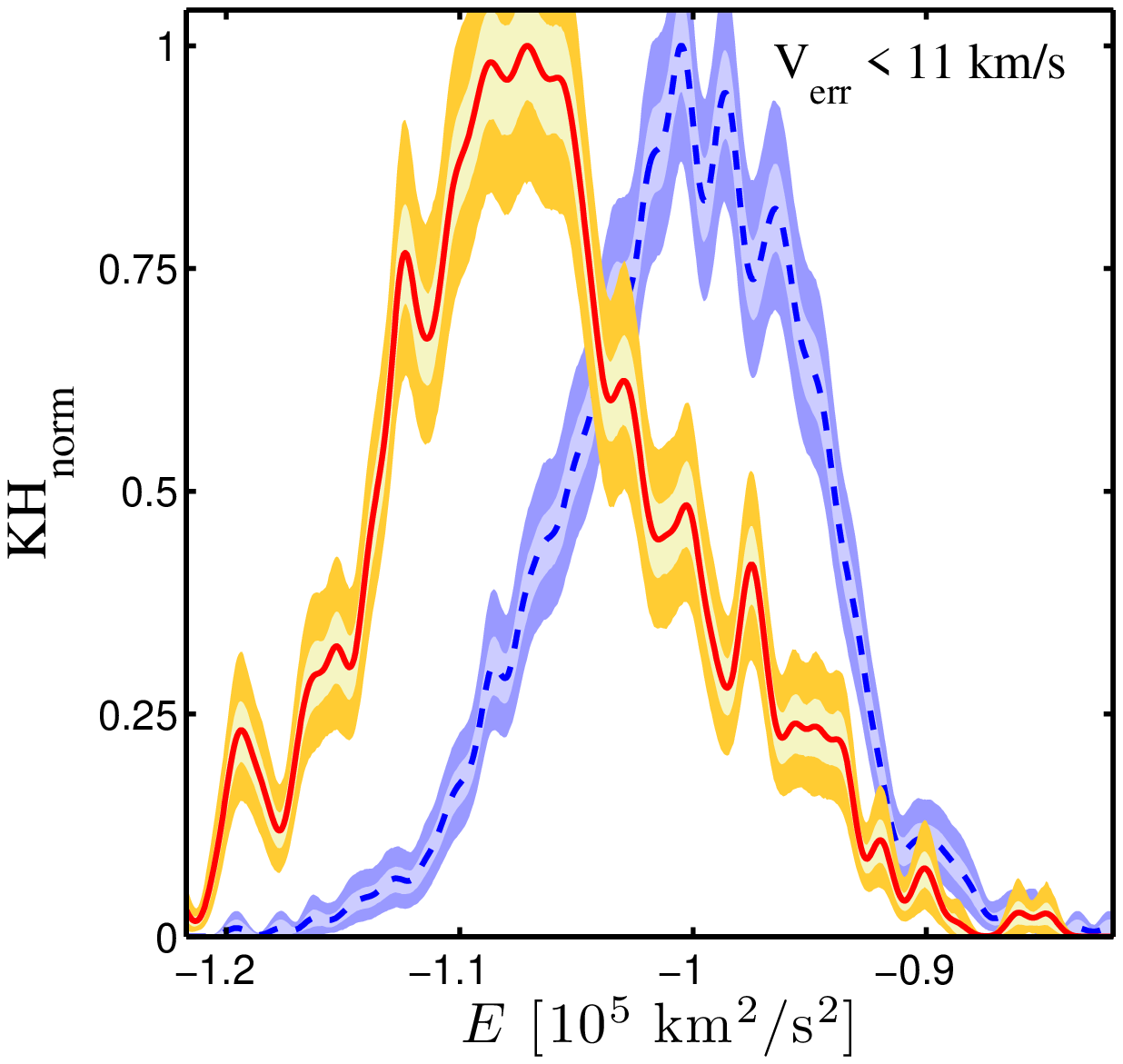}
\\
\includegraphics[width=82mm,clip]{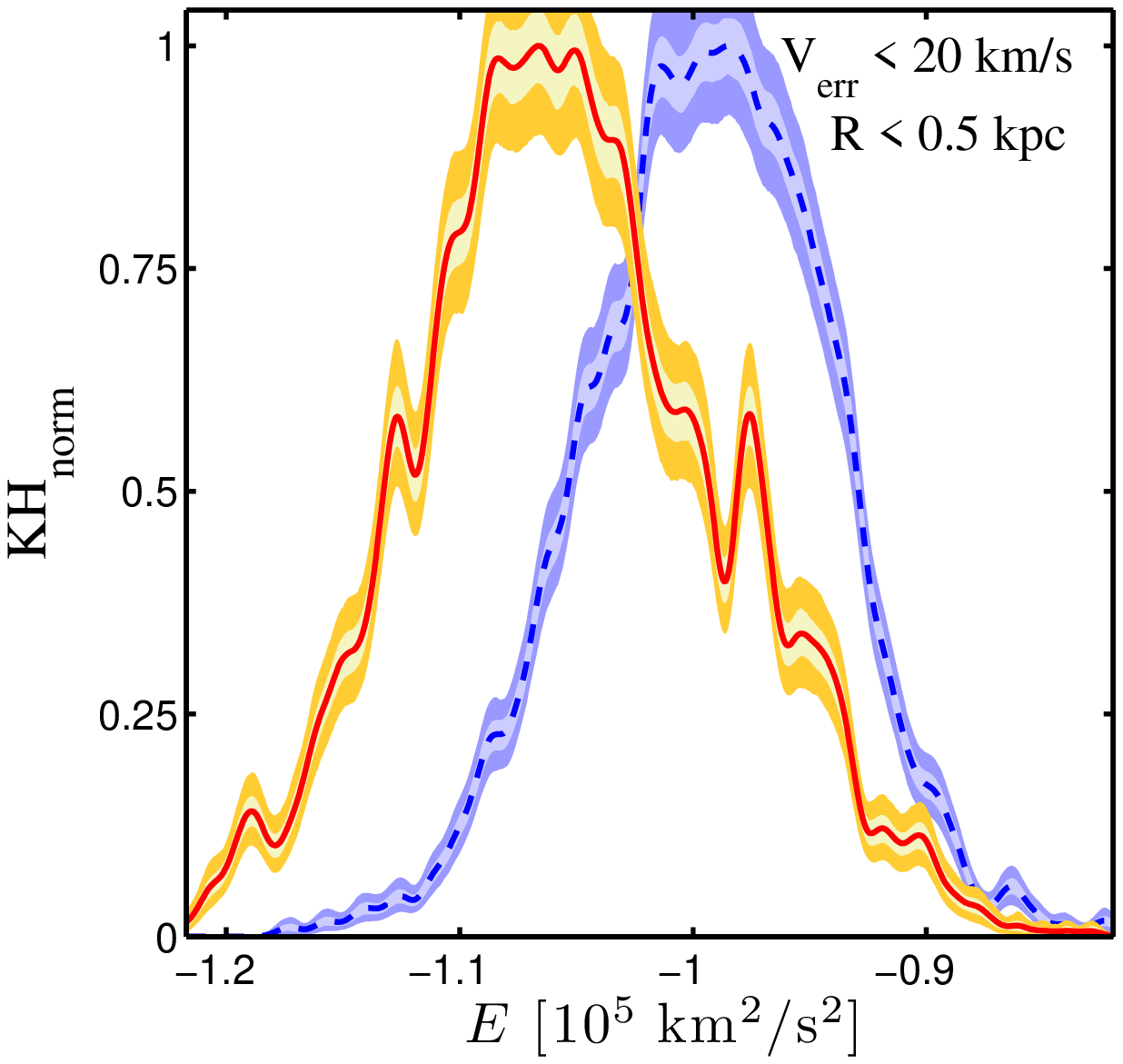}
\caption{Top panel:  Kernel histograms of energies  obtained from the
  SEGUE F/G-dwarf  thick disc (red solid  line) and the  thin disc (blue
  dashed line), when errors in  total velocity smaller than 11~km/s are
  considered. The shaded regions  indicate errors associated with poor
  sampling      of      the      underlying     distribution      (see
  Section~\ref{sec:th_tk}). Note that  most of the previously-observed
  features  in  Figure~\ref{fig:el_20} are  more  sharply defined,  as
  expected from  this more accurate  subsample. Note as well  that the
  thin disc  presents a smoother  distribution in $E_{\rm  norm}$ with
  respect to the thick disc. Bottom panel: As above, for stars located
  within $R \leq 0.5$ kpc and  total velocity errors smaller than 20~km/s.}
\label{fig:verr11}
\end{figure}

This exercise clearly indicates that, in order to detect signatures of
ringing in the  Milky Way disc, very accurate  6D phase-space catalogs
must be  explored. A highly accurate  subset of stars  can be obtained
from  our local  F/G-dwarf  sample, at  the  expense of  significantly
reducing  the  total  number  of  stars analysed.  The  top  panel  of
Figure~\ref{fig:verr11} shows the kernel  histograms of the thick disc
(red solid  line) and  the thin disc  (blue dashed line)  obtained when
only stars with  total velocity errors $V_{\rm err}  < 11$ km s$^{-1}$
are considered. As a result of  this more restrictive cut, we are left
with  $N_{\rm  stars}^{\rm  thick}   =  968$  thick-disc  and  $N_{\rm
  stars}^{\rm      thin}       =      1132$      thin-disc      stars,
respectively.  Interestingly,  not only  do  we  recover  most of  the
features   already  observed   in  Figure~\ref{fig:el_20},   but  also
structure here is more sharply  defined (as expected from our previous
discussion). Note, however, that  errors associated with poor sampling
of the underlying distribution are larger due to the smaller number of
stars.  This is  evident  by the  wider  areas covered  by the  shaded
regions, obtained  from the bootstrap  analysis of our  subsample (see
Section~\ref{sec:th_tk}). Nevertheless,  it is important  to note that
the  number  of thick-disc  stars  in  this  more accurate  subsample,
$N_{\rm stars}^{\rm  thick}$, is approximately equal to  the number of
particles analysed in the  $N$-body model, $N_{\rm part}^{\rm model}$,
shown in Figures~\ref{fig:model} and \ref{fig:mod_errors}.

Alternately, it is possible to obtain a more accurate subsample simply by
analysing a much more local subset of stars. In the bottom panel of
Figure~\ref{fig:verr11} we show the kernel histogram of $E$ obtained from a sample of
stars with heliocentric distances projected onto the plane $R \leq 0.5$
kpc. Note that by reducing the radius of our Solar neighbourhood sphere
we are not only constraining errors in the velocities of the stars, but
also in their positions. After the cut we obtain a thick-disc subsample of
$N_{\rm stars}^{\rm thick} = 2441$ stars, whereas the thin-disc subsample
contains $N_{\rm stars}^{\rm thin} = 3021$ stars. Comparison with the top
panel of Figure~\ref{fig:verr11} and Figure~\ref{fig:el_20} shows again
that many previously-observed features are preserved. Furthermore, some of
the features have become {\it even more significant}, in particular, the peak
located at $E \approx -0.97 \times 10^{5}$ km$^2$ s$^{-2}$. It is also
interesting to note that, as previously discussed in Section
\ref{sec:th_tk}, the thin-disc population exhibits  a smoother
energy distribution relative to that of the thick-disc population in both
of these panels. Notice that now both subsamples have comparable numbers of stars.

\begin{figure}
\includegraphics[width=82mm,clip]{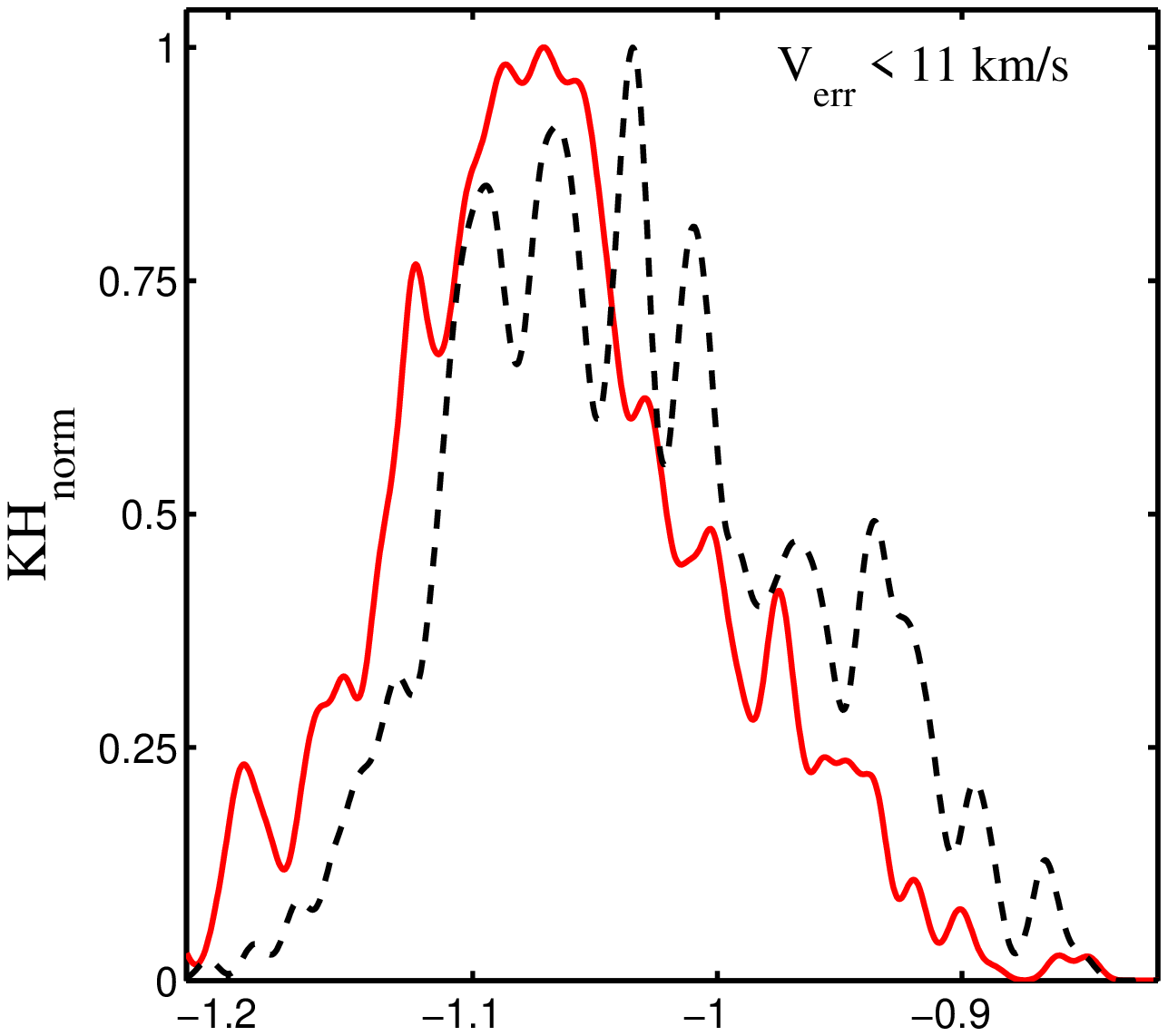}
\\
\includegraphics[width=82mm,clip]{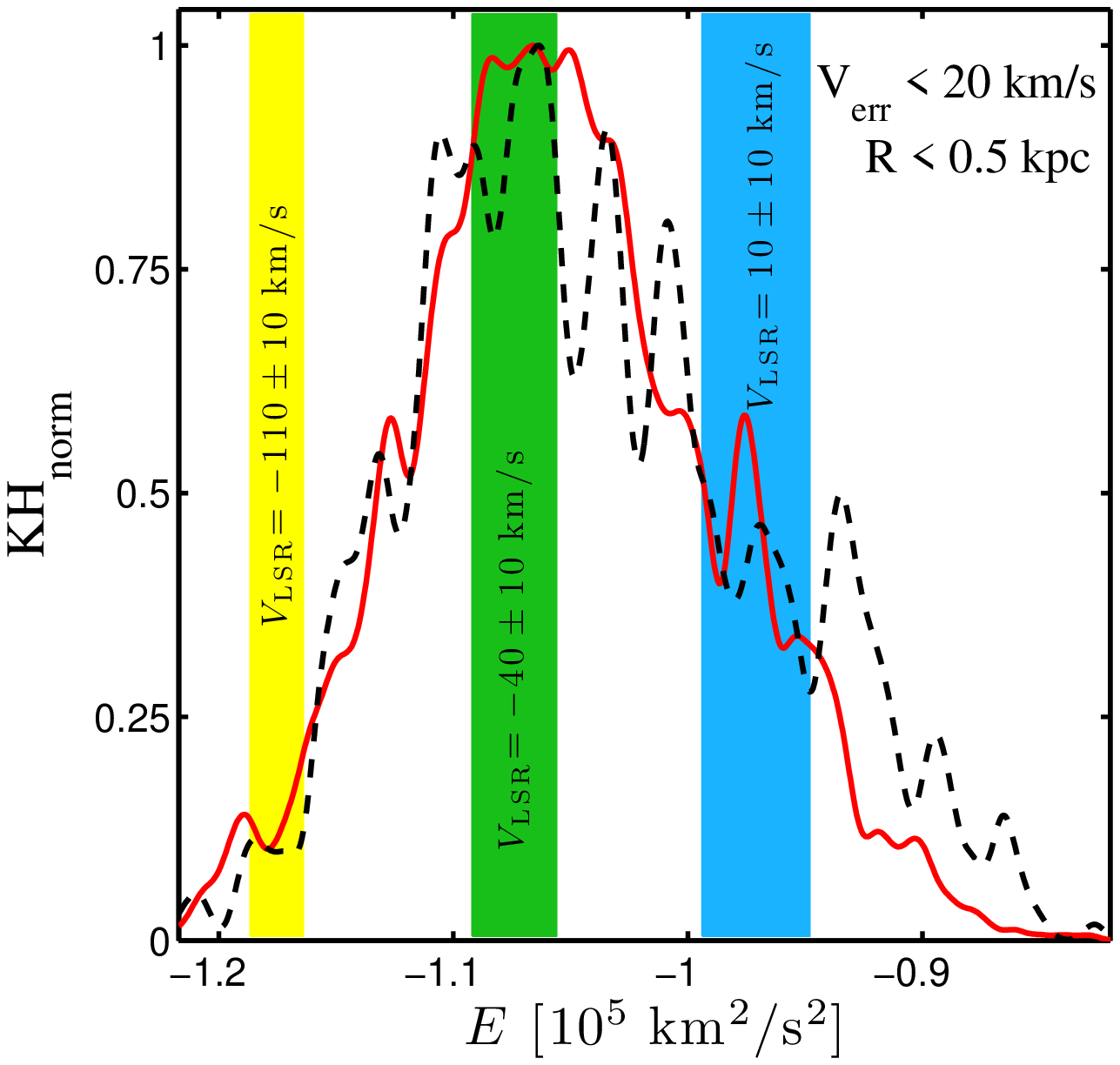}

\caption{Top panel:  Kernel histogram of  $E$ obtained from  the SCH06
  sample of  disc stars with total  velocity error $V_{\rm  err} < 11$
  km/s, shown with a black dashed line. The sample contains a total of
  603 stars.  Note  the large number of peaks  located at very similar
  values of $E$ as in the kernel histogram of the F/G-dwarf subsample,
  indicated with a red solid line.  The slight shift between these two
  distributions can  be accounted for by the  different volumes probed
  by  the  samples  (see  text),  as  well as  poor  sampling  of  the
  underlying energy  distribution. Bottom panel: As in  the top panel,
  for a sample of disc stars  with total velocity error $V_{\rm err} <
  20$  km/s.  This  larger sample  contains 813  stars.   The coloured
  areas indicate  the approximate energy range of  the Arcturus stream
  (yellow), the Hercules stream (green), and the most significant peak
  identified in the SEGUE F/G-dwarf sample (light blue).}
\label{fig:sch}
\end{figure}

\section{The Schuster et al. (2006) sample in $E-L_{z}$ space}
\label{sec:schuster_ener}

The previously-observed  peaks in the  kernel histograms of  our SEGUE
F/G-dwarf sample might  be an indication of ringing  in our own Galactic
thick disc. However, due to  the relatively small number of stars with
the  necessary small measurement  errors, uncertainties  associated with
poor sampling  of the underlying distribution are  (for some features)
still  large.  It  is  therefore  important  to  look  for  additional
evidences of  ringing in other  stellar catalogs already  available in
the  literature.  The  SCH06  is  particularly  well-suited  for  this
analysis, as it comprises a set  of old and metal-poor disc stars (see
Section~\ref{sec:schuster}).  Furthermore,  the  disc  stars  in  this
sample are all located within $d  < 0.2$ kpc, thus they probe a region
of the Solar neighbourhood that is different than that explored by the
more distant SEGUE F/G-dwarf sample.

The black  dashed line on  the top panel of  Figure~\ref{fig:sch} shows
the kernel  histogram obtained from  a subsample of likely  disc stars
from the  SCH06 catalog.  Following our previous  discussion, we first
consider stars with  total errors in velocities $V_{\rm  err} \leq 11$
km s$^{-1}$.  As a consequence,  we are left  with a total of  603 old
disc stars.  Interestingly, a large  number of peaks can  be observed,
most of them  located at values of $E$  consistent with those observed
in the kernel  histogram of F/G-dwarf subsample, shown  with a red solid
line.  As we explain  in Section~\ref{sec:sag_ps},  due to  the strong
radial and azimuthal  dependance on the location and  amplitude of the
peaks,  together  with the  poor  sampling  of  the underlying  energy
distribution, it  is not surprising  to find features in  these kernel
histograms slightly shifted with respect to one another.

The bottom  panel of  Figure~\ref{fig:sch} shows the  kernel histogram
from the  SCH06 catalog, obtained  after considering stars  with total
velocity errors  of $V_{\rm  err} \leq 20$  km s$^{-1}$.   This larger
sample  contains 813  stars.  Again,  note that  many of  the features
observed in  the KHs of both  the SCH06 and the  F/G-dwarf samples are
located at very similar values of  $E$. The colour-coded areas in
  this  panel  indicate the  approximate  energy  range  in which  the
  Arcturus and the Hercules moving groups are expected to be found. To
  compute  these energy ranges,  for simplicity,  we have  assumed for
  each moving group:

\begin{itemize}

\item  a   characteristic  $V$  velocity   component,  $V_{\rm  LSR}$
  \citep[see,   e.g.,][]{walter08,fux01,ant08,nhf,klement}.  Since  this
  characteristic   V   velocity  is   subject   to  relatively   large
  uncertainties, to  compute each energy  range we assumed $V  = V_{\rm
    LSR } \pm 10$ km s$^{-1}$

\item a characteristic $U$ velocity component equal to zero

\item a Galactocentric distance $R = 8$ kpc

\end{itemize}

The bottom  panel shows  that both moving  groups could  be associated
with  peaks in our  KH.  Note  that, as  previously explained  in this
Section, it is not surprising to find slight shifts in the location of
the  peaks between  the  two  analysed samples  due  to the  different
volumes  probed.  M09  considered  the possibility  that the  Arcturus
moving  group is the  result of  a perturbation  in the  Galactic disc
induced by a  minor merger event.  They concluded  that a minor merger
that took place $\sim 1.9$ Gyr ago could explain not only the presence
of  the Arcturus  moving  group, but  also  other previously  observed
features in  the local velocity field  \citep[see e.g.][]{ari_fu}.  It
is important to note that  streams with velocities larger than $V_{\rm
  LSR} \sim \pm  50$ km s$^{-1}$ are less likely to  be related to the
Galactic bar.  Interestingly, the short timescale associated with this
event is  consistent with  the estimated age  of the Galactic  bar, as
measured by \citet{cw}  ($< 3$ Gyr ago).  This  suggests that the same
event that could  have caused the formation of  the Galactic bar could
have also  left the  stellar disc unrelaxed,  thus giving rise  to the
observed high-velocity  stream.  On the other hand,  many studies have
successfully described the presence of  the Hercules moving group as a
result   of   a  resonant   interaction   with   the  bar   \citep[see
e.g.][]{walter,min07}.  However, as  shown by \citet{ant09,ant11}, the
effect that bars  and spiral density waves have  on the velocity field
of  galactic  discs   is  highly  degenerate\footnote{Note  that  both
  mechanisms  could   be  acting  simultaneously.}.    Therefore,  the
scenario in which the Hercules moving group was formed as the response
of the Galactic disc to a minor merger event cannot be ruled out.

\section{The Sagittarius dwarf galaxy as a possible perturber}

The impact that the Sagittarius dwarf galaxy may have had on the Galactic
disc has been previously considered by several authors. Recently, using
high-resolution $N$-body simulations, P11 showed that its gravitational
interaction may play a significant role on shaping the morphology of the
Galactic disc, by inducing the formation of rings, influencing the Galactic
bar, and flaring the outer disc. Its interaction could also explain the
formation of the Monoceros rings \citep{new02,quill09,md11,pur11}. However,
until now, its influence on the phase-space distribution of disc stars
located within the neighbourhood of the Sun has not been explored. In this
section we analyse the simulations presented by P11 to characterise the
effect of Sagittarius within this volume.

\begin{table}
\begin{minipage}{90mm} \centering
  \caption{Initial  properties  of   the  two  $N$-body  simulations
    analysed in Section 6.}
\label{table:Nmodel}
\begin{tabular}{@{}llllr} 
\hline 
\hline
Host & & & \\
\hline
DM halo & & & $N_{\rm part} = 2.65 \times 10^{7}$ \\
\hline 
Virial mass & $1 \times 10^{12}$ & & $[M_{\odot}]$ \\ 
Scale radius & $14.4$ & & [kpc] \\
\hline
Stellar disc & & & $N_{\rm part} = 3 \times 10^{6} $ \\
\hline
Mass & $3.59 \times 10^{10}$ & & [$M_{\odot}$] \\
Scale length & 2.84 & & [kpc] \\
Scale height & 0.43 & & [kpc] \\
\hline
Stellar bulge & & & $N_{\rm part} = 5 \times 10^{5} $ \\
\hline
Mass & $9.52 \times 10^{9}$ & & [$M_{\odot}$] \\
Effective radius & 0.56 & & [kpc] \\
S{\'e}rsic index & 1.28 & & \\
\hline
\hline
Satellites & & & \\
\hline
DM halo & {\it Light}  & {\it Heavy} &  $N_{\rm part} = 1.8 \times 10^{6}$ \\
\hline
Virial mass & $0.32 \times 10^{11}$ & $1 \times 10^{11}$  &[$M_{\odot}$] \\ 
Scale radius & $4.9$ & $6.5$ &  [kpc] \\
\hline  
Stellar spheroid & & &  $N_{\rm part} = 5 \times 10^{4} $ \\
\hline
Core radius & $1.5$ & $1.5$ & [kpc] \\
Tidal radius & $4$ & $4$ & [kpc] \\
Central vel. disp. & 23 &  30 & [km s$^{-1}$]
\end{tabular}
\end{minipage}
\end{table} 

\subsection{The simulations}

  Two  simulations with different models for  the Sagittarius dwarf
  galaxy progenitor were performed.  In both cases, the primary galaxy
  includes a  NFW dark matter  (DM) halo, an exponential  stellar disc
  and a central bulge following  a S{\'e}rsic profile.  The {\it Light
    Sgr}  and  {\it  Heavy  Sgr}  progenitors  were  self-consistently
  initialised  with a  NFW DM  halo and  a separate  stellar component
  following  a King  profile  \citep{king}.  Table  \ref{table:Nmodel}
  summarises  the numerical values  of the  parameters used  for these
  models.    Following   previous   work   on  the   Sgr   interaction
  \citep{kesel},  the  satellites were  launched  at  80~kpc from  the
  galactic centre in the plane  of the Milky Way, traveling vertically
  at 80~km  s$^{-1}$ toward the north galactic  pole.  The simulations
  reach a  present day configuration  after approximately 2.7  and 2.1
  Gyr of evolution, respectively.   Mass loss that would have occurred
  between  virial  radius  infall  and this  ``initial''  location  is
  accounted for by  truncating the progenitor DM halo  mass profile at
  the instantaneous Jacobi  tidal radius, $r_{\rm t} =  23.2$ and 30.6
  kpc, respectively.  This leaves a  total bound mass that is a factor
  of  $\sim 3$  smaller than  their effective  virial  mass originally
  assigned.  Each  of the model  Sgr progenitors experiences  two disc
  crossings, approaching  a third at  the present day.   The satellite
  first crosses the disc at a galactocentric distance of $\sim 20$ kpc
  approximately $\sim  1.75$ Gyr  ago, producing the  most significant
  perturbation.   The  progenitor loses  roughly  $75\%$  of its  dark
  matter mass (but little stellar  material) during this time.  In the
  bottom panel  of Figure~\ref{fig:xy_sag} we show  the time evolution
  of  the satellites'  galactocentric distance.   The initial  disc in
  both Sgr-infall models  is completely smooth at $t=0$  Gyr, and only
  develops a mild  bar when it is evolved in isolation  for a few Gyr.
  When compared to the isolated run, the {\it Light Sgr} model results
  in a  stronger bar  (see P11). Conversely,  as a result  of enhanced
  central  disc heating,  the  more massive  satellite suppresses  bar
  formation. The endstate  bar orientation, $\phi_{\rm bar}^{\rm sims}
  = 15^{o}  - 20^{o}$, is in  both cases consistent  with estimates of
  the long bar's orientation observed  in the center of the Milky Way,
  $\phi_{\rm  bar}^{\rm  MW} =  15^{o}  -  30^{o}$ \citep{biss}.   The
  simulations   used   the   parallel   $N$-body  tree   code   ChaNGa
  \citep{changa},  with  a   gravitational  softening  length  of  one
  parsec\footnote{ The  force softening  length is set  to roughly
      0.1 of the mean  disc interparticle spacing.}, and followed the
  evolution of 30 million particles with masses in the range 1.1 - 1.9
  $\times ~ 10^{4} {\rm M}_{\odot}$.

\subsection{The phase-space distribution of  Solar neighbourhood-like
  volumes}

\label{sec:sag_ps}

\begin{figure}
\begin{centering}
\hspace{-0.5cm}
\includegraphics[width=82mm,clip]{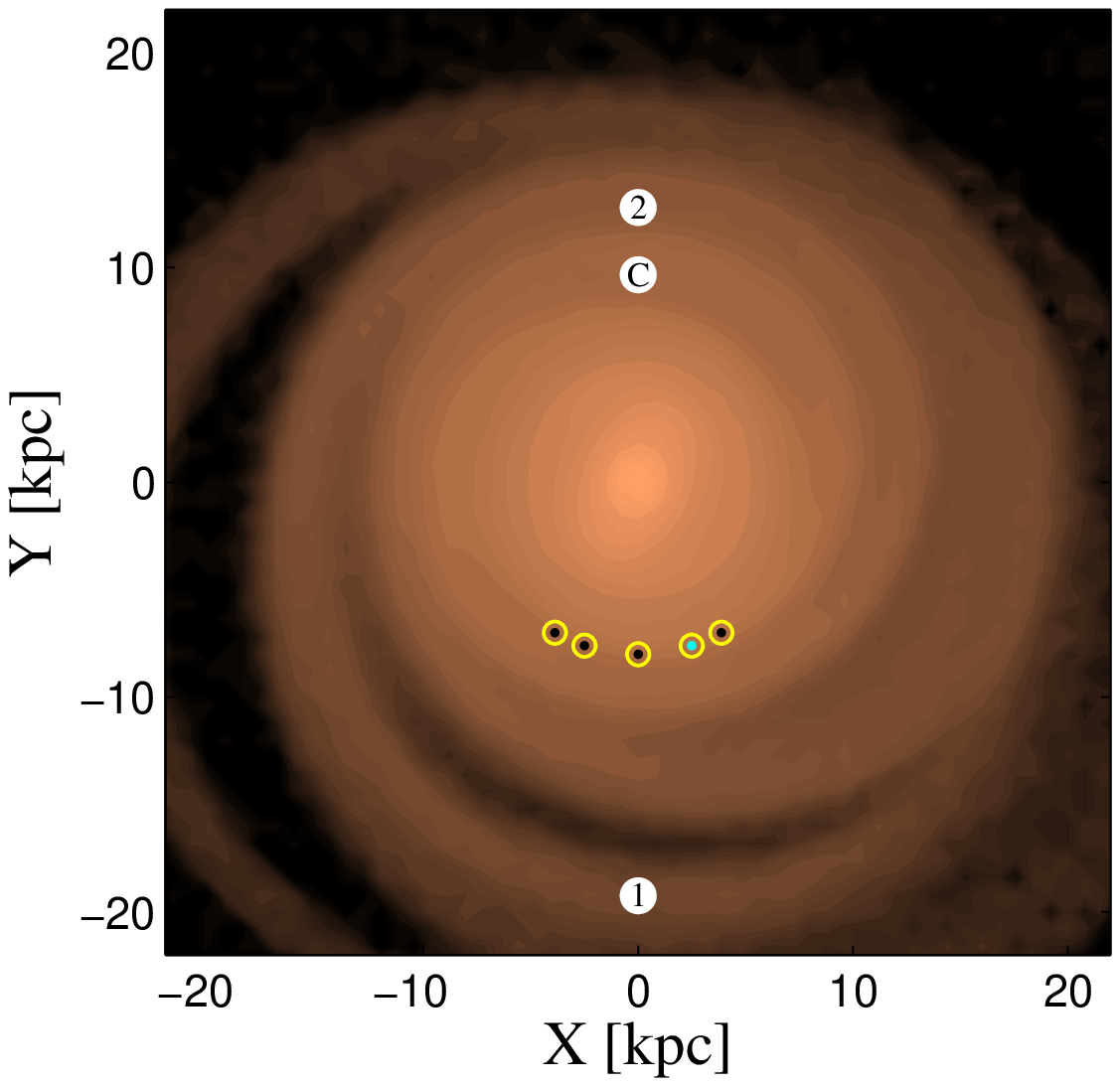}
\\
\includegraphics[width=79mm,clip]{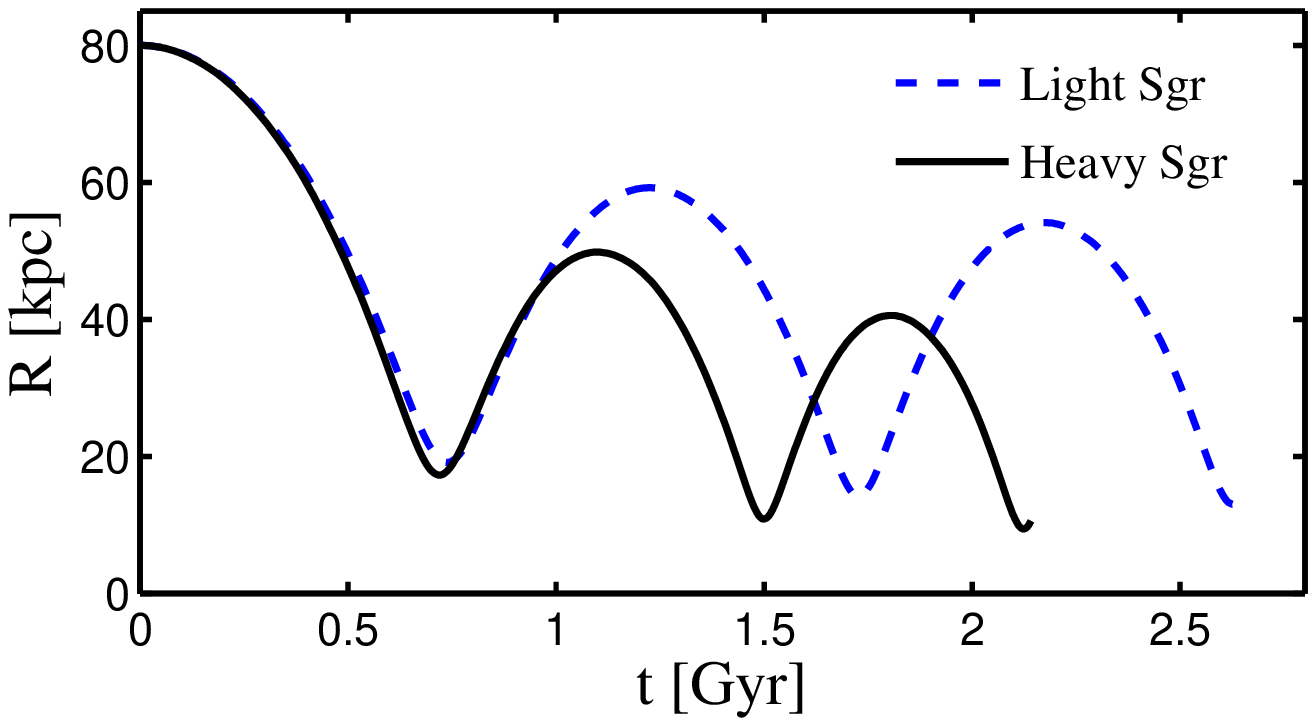}
\caption{Top panel: Distribution  of particles in the X  - Y plane, as
  obtained from the {\it Heavy Sgr} simulation after approximately 2.1
  Gyr  of  evolution.    The  different  colours  (contours)  indicate
  different   numbers  of   particles.   As   a  consequence   of  the
  non-axisymmetrical  energy kick  imparted by  the  satellite, spiral
  features  are induced  in the  galactic disc.   The location  of the
  Solar       neighbourhood-like       volumes       analysed       in
  Section~\ref{sec:sag_ps} are indicated with solar symbols.  The cyan
  dot indicates the local  volume that best resembles the observations
  in the  Solar Neighbourhood.  The white dots  labelled as  ``1'' and
  ``2'' trace the regions where  the {\it Heavy Sgr} crosses the plane
  of the disc for the first and second time, respectively. Its current
  projection on  the X - Y  plane is indicated with  the symbol ``C''.
  Bottom panel:  Evolution of the galactocentric distance  of both Sgr
  models as a function of time. }
\label{fig:xy_sag}
\end{centering}
\end{figure}

\begin{figure*}
\hspace{-0.16cm}
\includegraphics[width=176.5mm,clip]{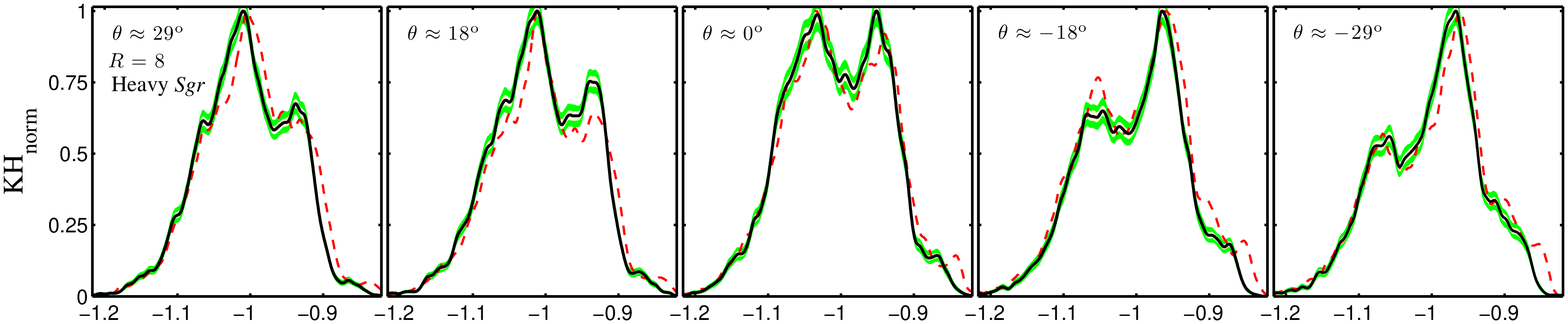}
\\
\includegraphics[width=175mm,clip]{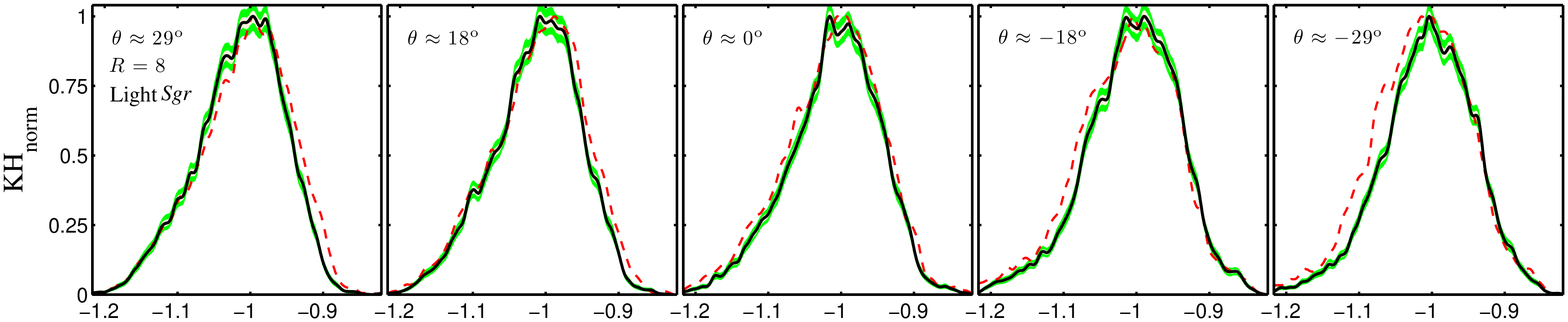}
\\
\includegraphics[width=176mm,clip]{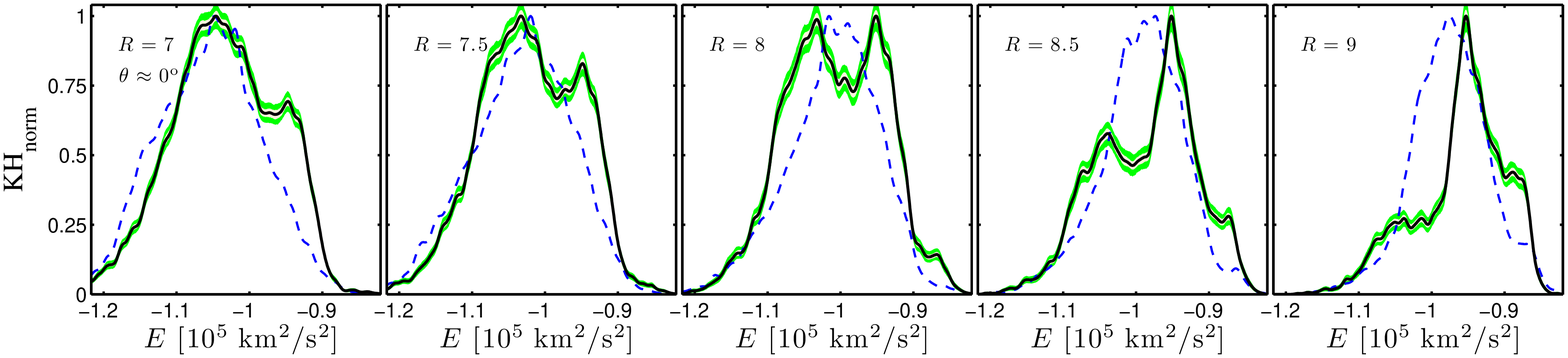}
\caption{Top panel:  Kernel histograms  of energies obtained  from the
  {\it Heavy  Sgr} simulation. Each  panel shows the  kernel histogram
  (KH)  obtained from  the distribution  of particles  within  a Solar
  neighbourhood-like  sphere  located  at a  different  galactocentric
  azimuthal angle,  $\theta$. The black solid lines  show KHs obtained
  after  computing  particle's  energy  using the  analytic  potential
  described  in Section~\ref{sec:gdwarf_ener}.  The  red dashed  lines
  show the  KHs obtained after computing particle's  energies from the
  $N$-body  potential. The shaded  regions indicate  errors associated
  with   poor   sampling   of   the   underlying   distribution   (see
  Section~\ref{sec:th_tk}).  The angle $\theta$ is measured clockwise,
  starting  from the  negative  Y axis  in Fig.~\ref{fig:xy_sag}.  The
  azimuthal  location of  the spheres  are indicated  in the  top left
  corner of each  panel. Note that all the  KHs exhibit multiple peaks
  associated with density  waves crossing the Solar neighbourhood-like
  volumes. Furthermore,  the locations of these  peaks strongly depend
  on  the azimuthal  location of  the sphere.  Note as  well  the good
  agreement between  the second panel  ($\theta \approx 18^{o}$  ) and
  the  KHs obtained for  our F/G-dwarf  thick-disc subsample  shown in
  Figures~\ref{fig:el_20}  and  \ref{fig:verr11}.  Middle  panels:  As
  above for the {\it Light Sgr} simulation. Note that, as expected for
  a less-massive  satellite, the peaks previously observed  on the KHs
  of the {\it Heavy Sgr} model  are either very mild or not present at
  all. Bottom panels: As above, now for spheres located at a different
  galactocentric radius, $R$. The  spheres are located at an azimuthal
  angles  $\theta =  0^{\rm o}$.  The galactocentric  distance  of the
  spheres  are indicated in  the top  left corner  of each  panel. The
  black solid  and blue  dashed lines show  the KHs obtained  from the
  from  the   {\it  Heavy  Sgr}  and  {\it   light  Sgr}  simulations,
  respectively. For the {\it  Heavy Sgr} simulation, the locations and
  amplitude  of the peaks  in the  KHs strongly  depend on  the radial
  position of the  spheres. Note as well the  similarities between the
  second panel ($R = 7.5$ kpc)  and the KHs obtained for our F/G-dwarf
  thick-disc  subsample.  For the  {\it  Light  Sgr} simulation,  mild
  signatures  of density  waves  can  be observed  only  in the  outer
  spheres,  where  the  perturbation  from  the  satellite  galaxy  is
  stronger.}
\label{fig:khh}
\end{figure*}

We now  characterise the phase-space distribution  of particles within
Solar neighbourhood-like volumes of 1~kpc radii. In these simulations,
according to  the location  of the Sagittarius  remnant and  its tidal
debris, the Sun should be located at approximately $(0,-8,0)$ kpc from
the  galactic centre.   In order  to examine  the dependencies  of the
particle's   phase-space  distribution   with  azimuthal   angle,  and
uncertainties  on the  location  of the  ``Sun''  associated with  the
model,  we  have  placed  5  spheres at  different  azimuthal  angles,
covering a  total of $\approx  58^{{\rm o}}$ around the  ``Sun.''  The
top panel of Figure~\ref{fig:xy_sag}  shows contour plots of the final
distribution of  particles in the X  - Y plane obtained  from the {\it
  Heavy Sgr} simulation.   The white dots trace the  regions where Sgr
crosses the plane  of the disc and its current projection  on the X- Y
plane. The locations of  our Solar neighbourhood spheres are indicated
with solar  symbols.  As shown  by P11, the  gravitational interaction
between  the  Sagittarius-like  satellite  and the  galactic  disc  is
sufficiently strong  to induce the formation of  transient spiral arms
or  rings.  These arms  can be  clearly observed  outside a  radius of
15~kpc, but not so clearly at  the Solar radius.  In the top panels of
Figure~\ref{fig:khh}  we show  the kernel  histograms of  $E$ obtained
from the  distribution of particles located within  these volumes. Let
us recall that throughout this work  we have kept fixed both the value
of $\Delta E$  and the value of the bandwidth  of the Gaussian kernel,
$\sigma$. The  black solid  lines show the  results obtained  when the
particle energies are computed using the analytic potential introduced
in  Section~\ref{sec:gdwarf_ener}, whereas  for the  red  dashed lines
energies were computed directly  from the $N$-body potential. The very
good agreement between the two KHs reflects the weak dependence of the
amount and distribution of features  with the particular choice of the
galactic potential.   Note that, for comparison, the  red dashed lines
have been slightly shifted in energy,  so that both KHs overlap in the
same energy  range. On average,  the number of particles  contained in
each sphere is $\approx 10,000$. It is interesting to observe that all
the analysed  volumes exhibit peaks  in their KHs; these  are features
associated  with  density waves.   We  see  strong  dependence of  the
distribution and  amount of peaks  with azimuthal angle,  $\theta$: as
$\theta$  is decreased,  power  from the  highest  amplitude peak  ($E
\approx  -1.01 \times  10^{5}$  km$^{2}$ s$^{-2}$  at $\theta  \approx
29^{o}$), or density wave, is  transferred towards a secondary peak at
higher energy  ($E \approx -0.96  \times 10^{5}$ km$^{2}$  s$^{-2}$ at
$\theta \approx -29^{o}$).  In addition,  the location of each peak is
shifted towards lower values of $E$ indicating that, on average, these
peaks are  populated by  more inner particles.   This is  the expected
behaviour from  a spiral pattern  traveling in this  angular direction
(see also Section  4.2 of G12).  Note the  similarities between the KH
in  the  second top  panel  of  Figure~\ref{fig:khh} ($\theta  \approx
18^{o}$)  with   the  KHs  obtained  from   our  F/G-dwarf  thick-disc
subsample,  shown  in  Figures~\ref{fig:el_20}  and  \ref{fig:verr11}.
This result  clearly indicates that  some of the features  observed in
the KH of the Milky Way  thick disc, especially the peak at $E \approx
-0.97 \times 10^{5}$ km$^2$ s$^{-2}$, could be associated with density
waves  excited   by  the  Sagittarius  dwarf   galaxy.   An  important
implication  of  the  strong  azimuthal  dependance  observed  in  our
simulations is  that the orbital  properties of the  Sagittarius dwarf
galaxy  could be  further  constrained by  fitting  the locations  and
amplitude of the significant peaks  in the models to those observed in
stellar samples.

In the  middle panels of  Figure~\ref{fig:khh} we show the  results of
the previous  analysis obtained  from the {\it  Light Sgr}  model. The
Solar neighbourhood-like spheres were placed at the same locations. As
before,  the black  solid lines  show the  kernel  histograms obtained
after  computing  particle  energies  using  the  analytic  potential,
whereas for the  red dashed lines the $N$-body  potential was used. As
expected from a less-massive  satellite (and thus a weaker perturber),
the peaks  previously observed  in the kernel  histograms of  the {\it
  Heavy Sgr} model  are either very mild or not present  at all. It is
interesting to note that, as previously discussed, the {\it Light Sgr}
model  results  in  a much  stronger  bar  than  the {\it  Heavy  Sgr}
model. The more massive satellite suppresses bar formation as a result
of enhanced central disc  heating. This clearly demonstrates that none
of the observed  features in these simulations could  be associated to
resonances  with the  galactic  bar.  Otherwise,  we  would expect  to
observe  higher-amplitude  features in  the  {\it  Light Sgr}  model's
kernel histograms  as a  result of the  interaction with  the stronger
bar. Thus,  models of the  Sagittarius dwarf galaxy progenitor  with a
total   mass  $\leq   10^{10.5}  {\rm   M}_{\odot}$  could   be  ruled
out. Moreover, note  that its total mass could  be further constrained
by comparing the  amplitudes of the significant peaks  observed in the
kernel histograms of models with those of the Milky Way disc.

\begin{figure*}
\hspace{-0.16cm}
\includegraphics[width=176.5mm,clip]{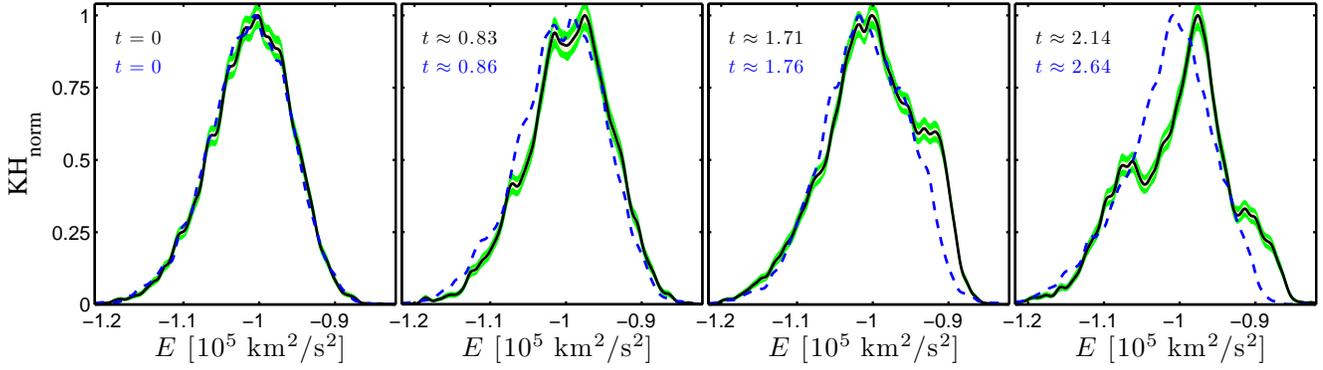}
\caption{  Kernel  histograms  of  energies  obtained  from  both  Sgr
  simulations at different times. Each  panel shows a KH obtained from
  a distribution of particles within a Solar neighbourhood-like sphere
  located at  $R = 8$ kpc  and rotating with an  angular frequency set
  set by  the velocity of  the LSR, after correcting  for axysimmetric
  drift.   The KH  are  shown  at four  different  times: the  initial
  conditions,  the  two  pericentre   passages  and  the  present  day
  configuration. The  black solid lines  show KHs obtained  {\it Heavy
    Sgr} model.  Note the smooth  distribution of particles  in energy
  space at  $t=0$.  Strong perturbation  in the disc  particles energy
  distribution are only excited  after the first pericentre passage of
  the satellite.  The  blue dashed line show the  KH obtained from the
  {\it  light  Sgr}  model.   Note  the  approximately  smooth  energy
  distribution at all times.}
\label{fig:kh_time}
\end{figure*}

In  the  bottom  panels   of  Figure~\ref{fig:khh}  we  explore  Solar
neighbourhood-like  spheres  of  1~kpc  radius  centred  at  different
galactocentric distances.   The spheres are separated by  0.5 kpc, and
are all  located at $\theta =  0^{\rm o}$. The black  solid lines show
the kernel histogram obtained from the {\it Heavy Sgr} model. A strong
dependence on  the amplitude of  each peak with  galactocentric radius
can  be observed. As  we move  towards the  galactic centre  we better
sample   inner  density   waves,   represented  by   peaks  at   lower
energies. Again, note the similarities between the KHs obtained in the
inner  spheres  (e.g.,  $R=7.5$  kpc)  and  those  obtained  from  our
F/G-dwarf sample  (see e.g.  Fig.~\ref{fig:verr11}).   The blue dashed
lines  in  the  bottom  panels  of Figure~\ref{fig:khh}  show  the  KH
obtained from the {\it Light  Sgr} model. As expected, mild signatures
of density waves can be observed  only in the outer spheres, where the
perturbation  from   the  satellite  galaxy  is   stronger.   In
  Figure~\ref{fig:kh_time} we  show the  time evolution of  the kernel
  histogram    of   particles   located    within   a    given   Solar
  neighborhood-like  sphere for  $ \approx  2.5$ Gyr.   To  track this
  volume over time,  our sphere is rotating with  an angular frequency
  set by the velocity of  the LSR, obtained after correcting the final
  mean rotational velocity profile for axysimmetric drift.  The KH are
  shown  at four  different  times: the  initial  conditions, the  two
  pericentre passages,  and the present day  configuration.  From this
  figure  it is  possible  to appreciate  the  smooth distribution  of
  particles in energy space at  $t=0$.  In addition, it is possible to
  observe  that  strong perturbations  in  the  disc particles  energy
  distribution are only excited  after the first pericentre passage of
  the {\it  Heavy Sgr}  satellite, as shown  by the black  solid line.
  Note that  the smaller number of peaks  in the KH of  the {\it Heavy
    Sgr} model with respect  to the KH shown in Figure~\ref{fig:model}
  is  due to  the  shorter period  of  time this  simulation has  been
  evolved. The blue dashed lines show KHs obtained from the {\it light
    Sgr} model. Notice the approximately smooth energy distribution at
  all times.

\section{Summary and Conclusions}
\label{sec:conclu}

We  have analysed the  SEGUE F/G-dwarf  and the  Schuster et  al. (2006)
stellar samples in  search for imprints of minor  merger events in the
Milky Way disc.  In contrast  to a  number of previous  studies, we  look for
perturbations in the phase-space of  the host disc rather than merger
remnants.  We  apply  the  same  analyses  to  high-resolution  N-body
simulations of  the interaction between  the Sagittarius dwarf galaxy  and the
Milky Way. We summarise our results as follows:
\begin{itemize}

\item The alpha-enhanced, likely thick-disc F/G-dwarf subsample exhibits
significant peaks in the stellar energy kernel histograms (KH). Conversely,
the thin-disc subsample shows a much smoother distribution. This is
consistent with the expectation that merger-induced perturbations in the
host galaxy disc (e.g., the Milky Way) create structure in the stellar phase-space
density, which survives until later times only in the old, hotter stellar
population.

\item  We show that, to unambiguously identify merger-induced waves
(ringing), samples of stars with velocity errors $V_{\rm err} < 11$ km
s$^{-1}$ are required. This depends on the time since the waves were
excited, as well as on the mass of the perturber.

\item Complementary evidence of ringing can be obtained from the SCH06
  catalog.  The energy  kernel histogram  of stars  here  also reveals
  peaks, most of them remarkably consistent with those observed in the
  F/G-dwarf sample.

\item  Simulations of the  interaction between the Milky  Way and
    the  Sagittarius  dwarf  galaxy  show  that  a  relatively-massive
    progenitor (i.e., masses of $\approx 10\%$ of the host) can induce
    density  waves in  the Galactic  disc.  Furthermore,  the features
    seen in the energy kernel  histogram of the observational data can
    be qualitatively well reproduced  with these simulations.  This is
    interesting since,  by matching features in  the kernel histograms
    of  both models  and observations,  one could  attempt  to further
    constrain  the orbital  properties of  Sagittarius, thanks  to the
    strong  azimuthal and  radial  dependance of  the distribution  of
    peaks.  Note,  however,  that  other possible  origins  for  these
    perturbations need to be ruled  out before performing this kind of
    analysis.

\item On  the contrary, low-mass  models of Sagittarius  barely excite
  density  waves in the  disc. Thus,  if we  assume that  the observed
  peaks in the kernel histogram of the F/G-dwarf sample are the response
  of the  Galactic disc  to the tidal  interaction with Sgr,  we could
  rule  out  such  low-mass   models.  Furthermore,  by  matching  the
  amplitudes of the peaks, one  could attempt to further constrain the
  total mass of the Sagittarius progenitor.

\end{itemize}

Comparing  Figures \ref{fig:verr11}  and \ref{fig:khh},  we  note that
variations  in the  azimuthal and  radial  positions of  the Sun  with
respect to  the Sagittarius orbital plane  can result in  similar structure in
the  stellar energy  distribution. This  degeneracy can  be  broken by
studying in detail the variation of features in the stellar phase-space 
distribution  when  position in  the  disc  is  varied. This  in  turn
requires  larger  and  deeper   stellar  samples,  which  will  become
available   in   ongoing   and   future  surveys,   such   as   APOGEE
\citep{apogee},  HERMES  \citep{hermes},  and  ultimately  {\it  Gaia}
\citep{gaia}.

Although  we cannot fully  discard the  possibility that  the observed
features  have   some  secular   origin,  the  much   smoother  energy
distribution of the thin-disc  population make this scenario unlikely.
If these  features had originated by resonant  interaction with, e.g.,
the Galactic  bar, we  would expect  them to be  stronger in  the thin
disc, rather than  in the thick disc.  This  result is consistent with
the  hypothesis  that features  associated  with  ringing should  last
longer in  the thick disc.   Thick-disc stars spend  relatively little
time near  the Galactic plane, where  heating from spiral  arms or the
bar and  scattering by  giant molecular clouds  is most  vigorous.  In
addition, features generated by the bar or spiral arms can be observed
as lumps in the u-v plane.   We have explored this plane, and found no
significant overdensities that could be linked to this kind of secular
perturbations.
 
In this study we considered  simulations with two different masses for
the  Sagittarius  progenitor, representing  a  lower  ($M_{\rm vir}  =
10^{10.5}~{\rm M}_{\odot}$)  and upper ($M_{\rm vir} =  10^{11} ~ {\rm
  M}_{\odot}$) limits.  We have found  that the {\it Light Sgr} cannot
explain the structure seen in  the energy distribution of our observed
samples. On the  other hand, the {\it Heavy  Sgr} model creates strong
disturbances in the  Milky Ways disc, resulting in  prominent peaks in
the energy kernel histogram in localised spatial volumes, perhaps even
stronger  than  what is  seen  in  the  observations (compare  Figures
\ref{fig:verr11} and \ref{fig:khh}). As  shown by G12, satellites with
masses $10\%$  and $20\%$ of the  mass of their  host generate density
waves, which can be followed for  no longer than 1 and 2~Gyr after the
satellite's full  disruption, respectively.  Close  encounters between
massive substructures and galactic  discs should be common occurrences
in a  $\Lambda$CDM cosmology since z  = 1 (Kazantzidis  et al.  2009).
However,  the short  relaxation time  scale involved  and the  lack of
observational evidence  of a merger  with such characteristics  in the
last 1-2 Gyr suggest that Sgr is the most likely perturber.

It is  important to  note that all  simulations analysed in  this work
considered initially stable, unperturbed  stellar discs. It remains to
be explored how a previously  perturbed disc, either by a merger event
or other dynamical process, would  react to the tidal interaction with
a  satellite  galaxy  such  as  Sgr. Future  studies  should  consider
simulations spanning a range of masses for the Sgr progenitor, between
the two  limits we  presented here, as  well as varying  Galactic disc
models,  which  can be  tuned  more finely  as  our  knowledge of  the
dynamics of the Milky Way expands.   In addition, it remains to be
  studied whether  perturbations from multiple  satellite galaxies can
  be successfully disentangled.

\section*{Acknowledgments}

We thank the anonymous referee for the useful comments and suggestions
that helped  to improve this work.  FAG was supported  through the NSF
Office of Cyberinfrastructure by grant PHY-0941373 and by the Michigan
State  University  Institute   for  Cyber-Enabled  Research.  BWO  was
supported in part  by the Department of Energy  through the Los Alamos
National  Laboratory Institute for  Geophysics and  Planetary Physics.
YSL and TCB  acknowledge partial support from grants  PHY 02-16783 and
PHY  08-22648: Physics  Frontiers Center/Joint  Institute  for Nuclear
Astrophysics (JINA), awarded by the U.S.  National Science Foundation.
DA acknowledges  support by the National Research  Foundation of Korea
to  the  Center  for  Galaxy  Evolution  Research.  \'AV  acknowledges
financial  support  from  the  European  Research  Council  under  the
European  Community’s Seventh Framework  Programme (FP7/2007-2013)/ERC
grant agreement n. 202781.

Funding for SDSS-III has been provided by the Alfred P. Sloan Foundation, the
Participating Institutions, the National Science Foundation, and the
U.S. Department of Energy. The SDSS-III web site is http://www.sdss3.org/.

SDSS-III is managed by the Astrophysical Research Consortium for the
Participating Institutions of the SDSS-III Collaboration, including the
University of Arizona, the Brazilian Participation Group, Brookhaven
National Laboratory, University of Cambridge, University of Florida, the
French Participation Group, the German Participation Group, the Instituto
de Astrofisica de Canarias, the Michigan State/Notre Dame/JINA
Participation Group, Johns Hopkins University, Lawrence Berkeley National
Laboratory, Max Planck Institute for Astrophysics, New Mexico State
University, New York University, the Ohio State University, University of
Portsmouth, Princeton University, University of Tokyo, the University of
Utah, Vanderbilt University, University of Virginia, University of
Washington, and Yale University.

\label{lastpage}
\end{document}